\documentclass[12pt]{article}
\usepackage{graphicx}
\begin{document}
%%----------------------------------------------------------------------------

\begin{center}

{\Large \bf Wigner's Space-time Symmetries based on the Two-by-two Matrices of
the Damped Harmonic Oscillators and the Poincar\'e Sphere}\\[10mm]

Sibel Ba{\c s}kal\\
Department of Physics, Middle East Technical University,
06800 Ankara, Turkey \\
e-mail: baskal@newton.physics.metu.edu.tr \\[5mm]

Young S. Kim \\
Center for Fundamental Physics, University of Maryland,\\ College Park,
Maryland 20742, U.S.A. \\e-mail: yskim@umd.edu   \\[5mm]

    Marilyn E. Noz \\[1mm]
Department of Radiology, New York University \\ New York, New York, 10016,
U.S.A. \\e-mail: marilyne.noz@gmail.com  \\[5mm]

\end{center}

\vspace{10mm}

\abstract{The second-order differential equation for a damped harmonic
oscillator can be converted to two coupled first-order equations, with two
two-by-two matrices leading to the group $Sp(2)$.  It is shown that this
oscillator system contains the essential features of Wigner's little groups
dictating the internal space-time symmetries of particles in the
Lorentz-covariant world.  The little groups are the subgroups of the
Lorentz group whose transformations leave the four-momentum of a given
particle invariant.  It is shown that the damping modes of the oscillator
correspond to the little groups for massive and imaginary-mass particles
respectively.  When the system makes the transition from  the oscillation
to damping mode, it corresponds to the little group for massless particles.
Rotations around the momentum leave the four-momentum invariant.  This
degree of freedom extends the $Sp(2)$ symmetry to that of $SL(2,c)$
corresponding to the Lorentz group applicable to the four-dimensional
Minkowski space. The Poincar\'e sphere contains the $SL(2,c)$ symmetry.
In addition, it  has a non-Lorentzian parameter allowing us to
reduce the mass continuously to zero.  It is thus possible to construct
the little group for massless particles from that of the massive particle
by reducing its mass to zero.  Spin-1/2 particles and spin-1 particles are
discussed in detail.}

\newpage

\section{Introduction}\label{intro}

We are quite familiar with the second-order differential equation
\begin{equation}\label{eq11}
m \frac{d^2y}{dt^2} + b \frac{d y}{dt} + K y = 0 ,
\end{equation}
for a damped harmonic oscillator. This equation has the same
mathematical form as
\begin{equation}\label{eq02}
L \frac{d^2Q}{dt^2} + R \frac{d Q}{dt} + \frac{1}{C} Q = 0 ,
\end{equation}
for electrical circuits, where $L, R,$ and $C$ are the inductance,
resistance, and capacitance respectively.  These two equations
play fundamental roles in physical and engineering sciences.
Since they start from the same set of mathematical equations, one
set of problems can be studied in terms of the other.  For instance,
many mechanical phenomena can be studied in terms of electrical
circuits.

\par
In Eq.(\ref{eq11}), when $b = 0$,
the equation is that of a simple harmonic oscillator with the frequency
$\omega = \sqrt{K/m}$.  As $b$ increases, the oscillation
becomes damped.  When $b$ is larger than $2\sqrt{Km}$, the
oscillation disappears, as the solution is a damping mode.

\par
Consider that increasing $b$ continuously, while difficult mechanically,
can be done electrically  using Eq.(\ref{eq02}) by adjusting
the resistance $R.$  The transition from the oscillation mode
to the damping mode is a continuous physical process.
\par

This $b$ term leads to energy dissipation, but is not regarded
as a fundamental force.  It is inconvenient in the Hamiltonian
formulation of mechanics and troublesome in transition
to quantum mechanics, yet, plays an important role
in classical mechanics.  In this paper
this term will help us understand the fundamental space-time symmetries of
elementary particles.

\par
We are interested in constructing the fundamental
symmetry group for particles in the Lorentz-covariant world.  For
this purpose, we transform the second-order differential
equation of Eq.(\ref{eq11}) to two coupled first-order equations
using two-by-two matrices.  Only two linearly
independent matrices are needed.  They are the anti-symmetric
and symmetric matrices
\begin{equation}\label{a11}
A =  \pmatrix{0 & -i \cr i & 0 },  \quad\mbox{and}\quad
S = \pmatrix{0 & i \cr i & 0} ,
\end{equation}
respectively. The anti-symmetric matrix $A$ is Hermitian and corresponds
to the oscillation part, while the symmetric $S$ matrix corresponds to
the damping.
\par

These two matrices lead to the $Sp(2)$ group consisting of two-by-two
unimodular matrices with real elements.  This group is isomorphic to
the three-dimensional Lorentz group applicable to two space-like and
one time-like coordinates.  This group is commonly called the $O(2,1)$
group.
\par
This $O(2,1)$ group can explain all the essential
features of Wigner's little groups dictating internal space-time
symmetries of particles~\cite{wig39}.  Wigner defined his little
groups as the subgroups of the Lorentz group whose transformations
leave the four-momentum of a given particle invariant.  He observed
that the little groups are different for massive, massless, and
imaginary-mass particles.  It has been a challenge to design a
mathematical model which will combine those three into one formalism,
but we show that the damped harmonic oscillator provides the
desired mathematical framework.
\par
For the two space-like coordinates, we can assign one of them
to the direction of the momentum, and the other to the direction
perpendicular to the momentum.  Let the direction of the momentum
be along the $z$ axis, and let the perpendicular direction be along
the $x$ axis.  We therefore study the kinematics of the group
within the $zx$ plane,  then see what happens when we rotate
the system around the $z$ axis without changing the
momentum~\cite{hks86jmp}.

\par
The Poincar\'e sphere for polarization optics contains the $SL(2,c)$
symmetry isomorphic to the four-dimensional Lorentz group applicable to
the Minkowski space~\cite{born80,hkn97,bross98,bk06jpa,kns13}.  Thus,
the Poincar\'e sphere extends Wigner's picture into the three space-like
and one time-like coordinates.  Specifically, this extension adds
rotations around the given momentum which leaves the four-momentum
invariant~\cite{hks86jmp}.

\par
While the particle mass is a Lorentz-invariant variable, the Poincar\'e
sphere contains an extra variable which allows the mass to change.
This variable allows us to take the mass-limit of the
symmetry operations.  The transverse rotational degrees of
freedom collapse into one gauge degree of freedom and
polarization of neutrinos is a consequence of the requirement of
gauge invariance~\cite{hks82,hks86ajp}.
\par
The $SL(2,c)$ group contains symmetries not seen in the three-dimensional
rotation group.  While we are familiar with two spinors for a spin-1/2
particle in nonrelativistic quantum mechanics, there are two
additional spinors due to the reflection properties of the Lorentz group.
There are thus sixteen bilinear combinations of those four spinors.
This leads to two scalars, two four-vectors, and one antisymmetric
four-by-four tensor.  The Maxwell-type electromagnetic field tensor
can be obtained as a massless limit of this tensor~\cite{bk97epl}.

\par
In Sec.~\ref{dampo}, we review the damped harmonic oscillator
in classical mechanics, and note that the solution can be either
in the oscillation mode or damping mode depending on the magnitude
of the damping parameter.  The translation of the second order equation
into a first order differential equation with two-by-two matrices is
possible.  This first-order equation is similar to the
Schr\"odinger equation for a spin-1/2 particle in a magnetic field.

\par

Section~\ref{sp2}  shows that the two-by-two matrices of Sec.~\ref{dampo}
can be formulated in terms of the $Sp(2)$ group.  These matrices can be
decomposed into the Bargmann and Wigner decompositions.   Furthermore, this
group is isomorphic to the three-dimensional Lorentz group with two space and
one time-like coordinates.
\par

In Sec.~\ref{internal}, it is noted that this three-dimensional Lorentz group
has all the essential features of Wigner's little groups which dictate the
internal space-time symmetries of the particles in the Lorentz-covariant
world.  Wigner's little groups are the subgroups of the Lorentz group whose
transformations leave the four-momentum of a given particle invariant.
The Bargmann Wigner decompositions are shown to be useful tools for studying
the little groups.
\par
In Sec.~\ref{complete}, we note that the given momentum is invariant under
rotations around it.  The addition of this rotational degree of freedom extends the
$Sp(2)$ symmetry to the six-parameter $SL(2,c)$ symmetry.  In the space-time
language, this extends the three dimensional group to the Lorentz group
applicable to three space and one time dimensions.

\par
Section~\ref{poincs} shows that the Poincar\'e sphere contains the symmetries
of $SL(2,c)$ group.  In addition, it contains an extra variable which allows
us to change the mass of the particle, which is not allowed in the Lorentz
group.
\par
In Sec.~\ref{massless}, the symmetries of massless particles are studied in
detail.  In addition to rotation around the momentum, Wigner's little group
generates gauge transformations.  While gauge transformations on spin-1
photons are well known, the gauge invariance leads to the polarization of
massless spin-1/2 particles, as observed in neutrino polarizations.

\par
In Sec.~\ref{tensor}, it is noted that there are four spinors for spin-1/2
particles in the Lorentz-covariant world.  It is thus possible to construct
sixteen bilinear forms, applicable to two scalars, and two vectors, and
one antisymmetric second-rank tensor.  The electromagnetic field tensor
is derived as the massless limit. This tensor is shown to be gauge-invariant.

\section{Classical Damped Oscillators}\label{dampo}

For convenience, we write Eq.(\ref{eq11}) as
\begin{equation}\label{eq22}
\frac{d^2y}{dt^2}  + 2\mu \frac{d y}{dt} + \omega^{2}  y = 0 ,
\end{equation}
with
\begin{equation}
\omega =  \sqrt{\frac{K}{m}}, \quad\mbox{and}\quad \mu = \frac{b}{2m} ,
\end{equation}
The damping parameter $\mu$ is positive when there are no external
forces. When $\omega$ is greater than $\mu$, the solution takes the form
\begin{equation}\label{dmp11a}
 y =  e^{-\mu t}\left[C_{1}\cos(\omega't) + C_{2} \sin(\omega' t)  \right] ,
\end{equation}
where
\begin{equation}\label{omega22}
\omega' = \sqrt{\omega^2 - \mu^2} ,
\end{equation}
and $C_{1}$ and $C_{2}$ are the constants to be determined by the
initial conditions.  This expression is for a damped harmonic
oscillator.  Conversely, when $\mu$ is greater than $\omega$, the
quantity inside the square-root sign is negative, then the solution becomes
\begin{equation}\label{dmp11b}
y = e^{-\mu t}\left[C_{3} \cosh(\mu' t) + C_{4} \sinh(\mu' t)\right] ,
\end{equation}
with
\begin{equation}\label{mu22}
\mu' = \sqrt{\mu^2 - \omega^2} ,
\end{equation}

\par
If $\omega = \mu$, both  Eq.(\ref{dmp11a}) and
Eq.(\ref{dmp11b})
collapse into one solution
\begin{equation}\label{dmp11c}
y(t) = e^{-\mu t}\left[C_{5} + C_{6}~t\right] .
\end{equation}

\par
These three different cases are treated separately in
textbooks.  Here we are interested in the transition
from Eq.(\ref{dmp11a}) to
Eq.(\ref{dmp11b}), via
Eq.(\ref{dmp11c}).  For convenience, we start from
 $\mu$ greater than $\omega$ with $\mu'$ given by
Eq.(\ref{mu22}).

\par
For a given value of $\mu$, the square root becomes zero when $\omega$
equals $\mu$.  If $\omega$ becomes larger, the square root
becomes imaginary and divides into two branches.
\begin{equation}
\pm i \sqrt{\omega^2 - \mu^2} .
\end{equation}
This is a continuous transition, but not an analytic continuation.
To study this in detail, we translate the second
order differential equation of Eq.(\ref{eq22}) into the first-order
equation with two-by-two matrices.
\par
Given the solutions of Eq.(\ref{dmp11a}),
and Eq.(\ref{dmp11c}), it is convenient to use $\psi(t)$
defined as
\begin{equation}
\psi(t) = e^{\mu t}y(t), \quad\mbox{and}\quad \quad y = e^{-\mu t}\psi(t) .
\end{equation}
Then $\psi(t)$ satisfies the differential equation
\begin{equation}
\frac{d^{2}\psi(t)}{d t^{2}} + (\omega^{2} - \mu^{2}) \psi(t) = 0.
\end{equation}
\par
\subsection{Two-by-two Matrix Formulation}

In order to convert this second order equation to a first order system
we introduce $\psi_{1,2}(t)=(\psi_{1}(t),\psi_{2}(t))$. Then we have
a system of two equations
\begin{eqnarray}\label{}
&{}& \frac{d \psi_{1}(t)}{d t} = (\mu - \omega)\psi_{2}(t), \nonumber
\\[1ex]
&{}& \frac{d \psi_{2}(t)}{d t} = (\mu + \omega)\psi_{1}(t)
\end{eqnarray}
which can be written in matrix form as
\begin{equation}\label{}
\frac{d}{dt}\pmatrix{\psi_{1} \cr \psi_{2}
}
=\pmatrix{0 & \mu -\omega \cr  \mu + \omega  & 0
}
\pmatrix{ \psi_{1} \cr \psi_{2}
}.
\end{equation}
\par

Using the Hermitian and anti-Hermitian matrices of
Eq.(\ref{a11}) in Sec.~\ref{intro}, we construct
the linear combination
\begin{equation} \label{ham11}
H = \omega \pmatrix{ 0 & -i \cr i & 0
}
 + \mu
\pmatrix{ 0 & i \cr i & 0}.
\end{equation}
We can then consider the first-order differential equation
\begin{equation}\label{eqn33}
i\frac{\partial}{\partial t} \psi(t) = H \psi(t).
\end{equation}
While this equation is like the Schr\"odinger equation for an electron in a
magnetic field, the two-by-two matrix is not Hermitian.  Its
first matrix is Hermitian, but the second matrix is anti-Hermitian.
It is of course an interesting problem to give a physical
interpretation to this non-Hermitian matrix in connection with
quantum dissipation~\cite{leggett87}, but this is beyond the scope
of the present paper.
The solution of Eq.(\ref{eqn33}) is
\begin{equation}\label{soln11}
 \psi(t) =
 \exp{\left\{
\pmatrix{ 0 & -\omega + \mu \cr \omega + \mu & 0
}
t\right\}}
\pmatrix{C_7 \cr C_8
},
\end{equation}
where $C_{7} = \psi_{1}(0)$ and  $C_{8} = \psi_{2}(0)$ respectively.
\par

\subsection{Transition from the Oscillation Mode to Damping Mode}

It appears straight-forward to compute this expression by a Taylor
expansion, but it is not.  This issue was extensively discussed in
previous papers by two of us~\cite{bk10jmo,bk13jmo}.  The key idea is
to write the matrix
\begin{equation}
\pmatrix{0 & -\omega + \mu \cr \omega + \mu & 0
}
\end{equation}
as a similarity transformation of
\begin{equation}
\omega' \pmatrix{0 & -1 \cr 1 & 0} \quad\quad
  (\omega > \mu ),
\end{equation}
and as that of
\begin{equation}
\mu' \pmatrix{ 0 & 1 \cr 1 & 0
} \quad\quad   (\mu  > \omega ),
\end{equation}
with $\omega'$ and $\mu'$
defined in Eq.(\ref{omega22}) and Eq.(\ref{mu22}), respectively.
\par
Then the Taylor expansion leads to
\begin{equation}\label{soln22a}
\pmatrix{\cos(\omega' t) &
   -\sqrt{(\omega - \mu)/(\omega + \mu)}~\sin(\omega' t) \cr
    \sqrt{(\omega + \mu)/(\omega - \mu)}~\sin(\omega' t) & \cos(\omega' t)
}    ,
\end{equation}
when $\omega$ is greater than $\mu$.  The solution $\psi(t)$ takes the form
\begin{equation}\label{soln22aa}
\pmatrix{ C_7 \cos(\omega' t) -
         C_8 \sqrt{(\omega - \mu)/(\omega + \mu)}~\sin(\omega' t) \cr
    C_7\sqrt{(\omega + \mu)/(\omega - \mu)}~\sin(\omega' t) +
    C_8 \cos(\omega' t)
}    .
\end{equation}
\par
If $\mu$ is greater than $\omega$, the Taylor expansion becomes
\begin{equation}\label{soln22b}
    \pmatrix{
    \cosh(\mu' t) &
    \sqrt{(\mu - \omega)/(\mu + \omega)}~\sinh(\mu't) \cr
    \sqrt{(\mu + \omega)/(\mu - \omega)}~\sinh(\mu't) & \cosh(\mu' t)}     .
\end{equation}
When $\omega$ is equal to $\mu$, both Eqs.(\ref{soln22a}) and
(\ref{soln22b}) become
\begin{equation}\label{soln22c}
\pmatrix{1  &  0 \cr  2 \omega t  & 1}  .
\end{equation}
\par
If $\omega$ is sufficiently close to but smaller than $\mu$, the matrix of
Eq.(\ref{soln22b}) becomes
\begin{equation}\label{m22b}
\pmatrix{ 1 + (\epsilon/2) (2\omega t)^2  &  + \epsilon (2\omega t)
   \cr   (2\omega t) &  1 + (\epsilon/2)(2\omega t)^2
}    ,
\end{equation}
with
\begin{equation}\label{epsil}
\epsilon = \frac{\mu - \omega}{\mu + \omega} .
\end{equation}
If $\omega$ is sufficiently close to $\mu$, we can let
\begin{equation}
\mu + \omega = 2\omega , \quad\mbox{and}\quad
 \mu - \omega = 2\mu \epsilon .
\end{equation}
If $\omega$ is greater than $\mu$, $\epsilon$ defined in Eq.(\ref{epsil})
becomes negative, the matrix of Eq.(\ref{soln22a}) becomes
\begin{equation}\label{m22a}
\pmatrix{ 1 - (-\epsilon/2) (2\omega t)^2  &  - (-\epsilon) (2\omega t)
   \cr  2\omega t  &  1 - (-\epsilon/2)(2\omega t)^2
}.
\end{equation}

We can rewrite this matrix as
\begin{equation}\label{m22aa}
\pmatrix{1 - (1/2)\left[\left(2\omega\sqrt{-\epsilon}\right) t\right]^2 &
      - \sqrt{-\epsilon} \left[\left(2\omega\sqrt{-\epsilon}\right) t\right]
\cr
    2\omega t & 1 - (1/2)\left[\left(2\omega\sqrt{-\epsilon}\right)
t\right]^2
} .
\end{equation}
If $\epsilon$ becomes positive, Eq.(\ref{m22b}) can be written as
\begin{equation}\label{m22bb}
\pmatrix{1 + (1/2)\left[\left(2\omega\sqrt{\epsilon}\right) t\right]^2 &
      \sqrt{\epsilon} \left[\left(2\omega\sqrt{\epsilon}\right) t\right] \cr
    2\omega t & 1 + (1/2)\left[\left(2\omega\sqrt{\epsilon}\right)t\right]^2} .
\end{equation}

\begin{figure}%[thb]
\centerline{\includegraphics[scale=0.7]{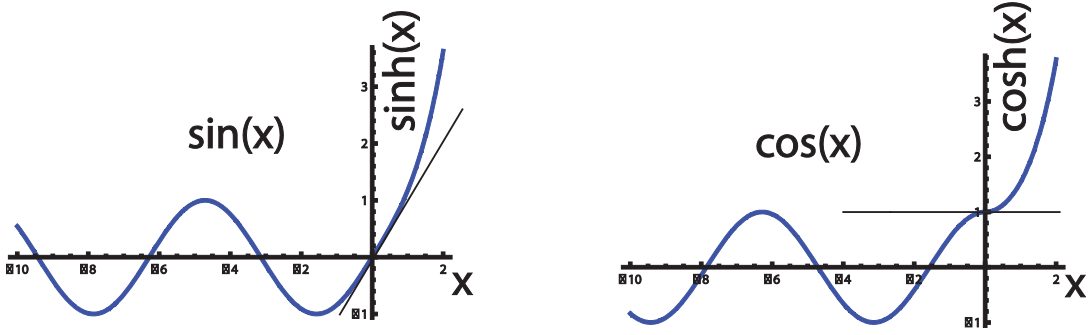}}
\caption{Transitions from $\sin$ to $\sinh$, and from $\cos$ to $\cosh.$
They are continuous transitions.  Their first derivatives are also
continuous, but the second derivatives are not.  Thus, they are not
analytically but only tangentially continuous.}\label{sincos}
\end{figure}

The transition from Eq.(\ref{m22aa}) to Eq.(\ref{m22bb}) is continuous as
they become identical when $\epsilon = 0.$  As $\epsilon$ changes its sign, the
diagonal elements of above matrices tell us how $\cos(\omega' t) $ becomes
$\cosh(\mu' t)$.  As for the upper-right element element, $- \sin(\omega't)$
becomes $\sinh(\mu' t).$  This non-analytic continuity is illustrated in
Fig.~\ref{sincos}.
\par

%----------------------------------------------------------------------
%----------------------------------------------------------------------

\subsection{Mathematical Forms of the Solutions}\label{mform}

In this section, we use the Heisenberg approach to the problem, and obtain
the solutions in the form of two-by-two matrices.  We note that

\begin{itemize}

\item[1.] For the oscillation mode, the trace of the matrix is smaller than 2.
The solution takes the form of
\begin{equation}\label{trace01}
\pmatrix{\cos(x)  & - e^{-\eta} \sin(x) \cr e^{\eta}\sin(x) & \cos(x)},
 \end{equation}
with trace $2\cos(x)$.  The trace is independent of $\eta$.
\item[2.] For the damping mode, the trace of the matrix is greater than 2.
\begin{equation}\label{trace02}
\pmatrix{\cosh(x)  & e^{-\eta} \sinh(x) \cr e^{\eta}\sinh(x) & \cosh(x)},
 \end{equation}
with trace $2\cosh(x)$.  Again, the trace is independent of $\eta$.
\item[3.] For the transition mode, the trace is equal to 2, and the matrix is
triangular and takes the form of
\begin{equation}\label{trace03}
\pmatrix{1 & 0 \cr \gamma & 1} .
\end{equation}

\end{itemize}
\par

When $x$ approaches zero, the  Eq.(\ref{trace01}) and Eq.(\ref{trace02}) take the form
\begin{equation}\label{trace11}
\pmatrix{1 - x^2/2 & -x e^{-\eta} \cr x e^{\eta} & 1 - x^2/2}, \quad\mbox{and}\quad
\pmatrix{1 + x^2/2 & x e^{-\eta} \cr x e^{\eta} & 1 + x^2/2},
\end{equation}
respectively.
These two matrices have the same lower-left element.  Let us
fix this element to be a positive number $\gamma$.  Then
\begin{equation}
x =  \gamma e^{-\eta} .
\end{equation}
Then the matrices of Eq.(\ref{trace11}) become
\begin{equation}\label{trace12}
\pmatrix{1 - \gamma^2 e^{-2\eta} /2 & -\gamma e^{-2\eta} \cr \gamma & 1 - \gamma^2 e^{-2\eta} /2},
\quad\mbox{and}\quad
\pmatrix{1 + \gamma^2 e^{-2\eta} /2 & \gamma e^{-2\eta} \cr \gamma & 1 + \gamma^2 e^{-2\eta} /2},
\end{equation}
If we introduce a small number $\epsilon$
defined as
\begin{equation}
\epsilon = \sqrt{\gamma} e ^{-\eta} ,
\end{equation}
the matrices of Eq.(\ref{trace12}) become
\begin{equation}\label{trace13}
\begin{array}{cc}
\pmatrix{e^{-\eta/2} & 0 \cr 0 & e^{\eta/2}}
\pmatrix{1 - \gamma \epsilon^2  /2 & \sqrt{\gamma}\epsilon \cr
\sqrt{\gamma}\epsilon & 1 - \gamma \epsilon^2/2 }
\pmatrix{e^{\eta/2} & 0 \cr 0 & e^{-\eta/2}} ,\\ \\
\pmatrix{e^{-\eta/2} & 0 \cr 0 & e^{\eta/2}}
\pmatrix{1 + \gamma \epsilon^2/2  & \sqrt{\gamma}\epsilon \cr
\sqrt{\gamma}\epsilon & 1 + \gamma \epsilon^2/2}
\pmatrix{e^{\eta/2} & 0 \cr 0 & e^{-\eta/2}}
\end{array}
\end{equation}
respectively,
with $e^{-\eta} = \epsilon/\sqrt{\gamma} $.

\section{Groups of Two-by-two Matrices}\label{sp2}

If a two-by-two matrix has four complex elements, it has eight independent
parameters. If the determinant of this matrix is one, it is known as an
unimodular matrix and the number of independent
parameters is reduced to six.
The group of two-by-two unimodular matrices is called $SL(2,c)$.
This six-parameter group is isomorphic to the Lorentz group applicable to
the Minkowski space of three space-like and one time-like
dimensions~\cite{knp86}.
\par
We can start with two subgroups of $SL(2,c)$.
\begin{itemize}
\item[1.] While the matrices of $SL(2,c)$ are not unitary, we can consider
   the subset consisting of unitary matrices.  This subgroup is called
   $SU(2)$, and is isomorphic to the three-dimensional rotation group.
   This three-parameter group is the basic scientific language for
   spin-1/2 particles.

\item[2.] We can also consider the subset of matrices with real elements.
   This three-parameter group is called $Sp(2)$ and is isomorphic to the
   three-dimensional Lorentz group applicable to two space-like and one
   time-like coordinates.
\end{itemize}
\par

In the Lorentz group, there are three space-like dimensions with $x, y,$
and  $z$ coordinates.  However, for many physical problems, it is more
convenient to study the problem in the two-dimensional $(x, z)$ plane first
and generalize it to three-dimensional space by rotating the system around
the $z$ axis.  This process can be called Euler decomposition and
Euler generalization~\cite{hks86jmp}.
\par

\par
First we study $Sp(2)$ symmetry in detail, and
achieve the generalization by augmenting the two-by-two matrix corresponding
to the rotation around the $z$ axis.
In this section, we study in detail properties of $Sp(2)$ matrices,
then generalize them to $SL(2,c)$ in Sec.~\ref{complete}.
\par
There are three classes of $Sp(2)$ matrices.  Their traces can be
smaller or greater than two, or equal to two.  While
these subjects are already discussed in the literature~\cite{barg47,iwa49,guil84}
our main interest is what happens as the trace goes from less than
two to greater than two.   Here we are guided by the model we have discussed in Sec.~\ref{dampo},
which accounts for the transition from the oscillation mode to the damping mode.

\subsection{Lie Algebra of Sp(2)}

The two linearly independent matrices of Eq.(\ref{a11}) can be written as
\begin{equation}\label{gen11}
K_1 = \frac{1}{2}\pmatrix{ 0 & i \cr i & 0} , \quad\mbox{and}\quad
J_2 = \frac{1}{2}\pmatrix{ 0 & -i \cr i & 0} .
\end{equation}
However, the Taylor series expansion of the exponential form of
Eq.(\ref{soln22a}) or Eq.(\ref{soln22b}) requires an additional matrix
\begin{equation}\label{gen66}
K_3 = \frac{1}{2}
\pmatrix{ i & 0 \cr 0 & -i} .
\end{equation}
These matrices satisfy the following closed set of commutation
relations.
\begin{equation}\label{sp2lie11}
 \left[K_1, J_2\right] = i K_3, \qquad \left[J_2, K_3\right] = i K_1, \qquad
\left[K_3, K_1\right] = -i J_2 .
\end{equation}
These commutation relations remain invariant under Hermitian conjugation,
even though $K_1$ and $K_3$ are anti-Hermitian.  The algebra generated by
these three matrices is known in the literature as the group
$Sp(2)$~\cite{guil84}.
Furthermore, the closed set of commutation relations is commonly called the
Lie algebra.  Indeed, Eq.(\ref{sp2lie11}) is the Lie algebra of the $Sp(2)$
group.
\par

The Hermitian matrix $J_2$ generates the rotation matrix
\begin{equation}\label{roty22}
R(\theta) = \exp{\left(-i\theta J_{2}\right)} =
\pmatrix{ \cos(\theta/2) & -\sin(\theta/2)
       \cr \sin(\theta/2) & \cos(\theta/2)},
\end{equation}
and the anti-Hermitian matrices $K_1$ and $K_2$, generate the following
squeeze matrices.
\begin{equation}\label{boostx22}
 S(\lambda) = \exp{\left(-i\lambda K_{1}\right)}
 =  \pmatrix{\cosh(\lambda/2) & \sinh(\lambda/2)
       \cr \sinh(\lambda/2) & \cosh(\lambda/2)},
\end{equation}
and
\begin{equation}\label{boostz22}
   B(\eta) = \exp{\left(-i\eta K_{3}\right)}
   = \pmatrix{
\exp{(\eta/2)} & 0 \cr 0 & \exp{(-\eta/2)}} ,
\end{equation}
respectively.
\par

Returning to the Lie algebra of Eq.(\ref{sp2lie11}),  since $K_1$ and
$K_3$ are anti-Hermitian, and $J_2$ is Hermitian, the set of commutation
relation is invariant under the Hermitian conjugation.  In other words,
the commutation relations remain invariant, even  if we change the sign
of $K_1$ and $K_3$, while keeping that of $J_2$ invariant.
Next, let us take the complex conjugate of the entire system.  Then both
the $ J$ and $K$ matrices change their signs.

\subsection{Bargmann and Wigner Decompositions}\label{bwdecom}

Since the $Sp(2)$ matrix has three independent parameters, it can be written
as~\cite{barg47}
\begin{equation}\label{bwd01}
\pmatrix{\cos\left(\alpha_1/2\right) & -\sin\left(\alpha_1/2\right)  \cr \sin\left(\alpha_1/2\right) & \cos\left(\alpha_1/2\right)}
\pmatrix{\cosh\chi & \sinh\chi  \cr \sinh\chi & \cosh\chi}
\pmatrix{\cos\left(\alpha_2/2\right) & -\sin\left(\alpha_2/2\right)  \cr \sin\left(\alpha_2/2\right) & \cos\left(\alpha_2/2\right)}
\end{equation}
This matrix can be written as
\begin{equation}\label{bwd02}
\pmatrix{\cos(\delta/2) & -\sin(\delta/2)  \cr \sin(\delta/2) & \cos(\delta/2)}
\pmatrix{a & b \cr c & d}
\pmatrix{\cos(\delta/2) & \sin(\delta/2)  \cr -\sin(\delta/2) & \cos(\delta/2)}
\end{equation}
where
\begin{equation}\label{bwd03}
\pmatrix{a & b \cr c & d} =
\pmatrix{\cos(\alpha/2) & -\sin(\alpha/2)  \cr \sin(\alpha/2) & \cos(\alpha/2)}
\pmatrix{\cosh\chi & \sinh\chi  \cr \sinh\chi & \cosh\chi}
\pmatrix{\cos(\alpha/2) & -\sin(\alpha/2)  \cr \sin(\alpha/2) & \cos(\alpha/2)}
\end{equation}
 with
\begin{equation}
\delta = \frac{1}{2}\left(\alpha_1 - \alpha_2\right) , \quad\mbox{and}\quad
\alpha = \frac{1}{2}\left(\alpha_1 + \alpha_2\right)  .
\end{equation}
If we complete the matrix multiplication of Eq.(\ref{bwd03}), the result is
\begin{equation}\label{bwd06}
\pmatrix{(\cosh\chi)\cos\alpha & \sinh\chi - (\cosh\chi)\sin\alpha  \cr
 \sinh\chi + (\cosh\chi)\sin\alpha  & (\cosh\chi)\cos\alpha } .
\end{equation}
We shall call hereafter the decomposition of Eq.(\ref{bwd03}) the Bargmann decomposition.
This means that every matrix in the $Sp(2)$ group can be brought to the Bargmann
decomposition by a similarity transformation of rotation, as given in
Eq.(\ref{bwd02}).  This decomposition leads to an equidiagonal matrix with
two independent parameters.
\par
For the matrix of Eq.(\ref{bwd03}), we can now consider the following three cases.
Let us assume that $\chi$ is positive, and the angle $\theta$ is less than
$90^o$.  Let us look at the upper-right element.
\begin{itemize}
\item[1.] If it is negative with $[\sinh\chi < (\cosh\chi)\sin\alpha]$, then the trace
of the matrix is  smaller than 2, and the matrix can be written as
\begin{equation} \label{bwd11}
\pmatrix{\cos(\theta/2) & - e^{-\eta}\sin(\theta/2)  \cr  e^{\eta}\sin(\theta/2) & \cos(\theta/2)},
\end{equation}
with
\begin{equation} \label{bwd12}
\cos(\theta/2) = (\cosh\chi) \cos\alpha, \quad\mbox{and}\quad e^{-2\eta} =
\frac{(\cosh\chi)\sin\alpha - \sinh\chi}{(\cosh\chi)\sin\alpha  + \sinh\chi}.
\end{equation}
\item[2.] If it is positive with $[\sinh\chi >
  (\cosh\chi)\sin\alpha)]$, then the trace is
greater than 2, and the matrix can be written as
\begin{equation}
\pmatrix{\cosh(\lambda/2) & e^{-\eta}\sinh(\lambda/2)  \cr
   e^{\eta}\sinh(\lambda/2) & \cosh(\lambda/2)},
\end{equation}
with
\begin{equation}
\cosh(\lambda/2) = (\cosh\chi) \cos\alpha, \quad\mbox{and}\quad e^{-2\eta} =
\frac{\sinh\chi - (\cosh\chi)\sin\alpha}{(\cosh\chi)\sin\alpha  + \sinh\chi}.
\end{equation}
\item[3.] If it is zero with $[(\sinh\chi = (\cosh\chi)\sin\alpha)]$,
  then the trace is
equal to 2, and the matrix takes the form
\begin{equation}
\pmatrix{1 & 0 \cr 2\sinh\chi & 1} ,
\end{equation}
\end{itemize}
The above repeats the mathematics given in
Subsec.~\ref{mform}.

\par
Returning to Eq.(\ref{bwd11}) and Eq.(\ref{bwd12}),  they can be decomposed
into
\begin{equation} \label{bwd16}
M(\theta,\eta) = \pmatrix{e^{\eta/2} & 0  \cr 0 & e^{-\eta/2}}
\pmatrix{\cos(\theta/2) & - \sin(\theta/2)  \cr \sin(\theta/2) & \cos(\theta/2)}
\pmatrix{e^{-\eta/2} & 0  \cr 0 & e^{\eta/2}} ,
\end{equation}
and
\begin{equation} \label{bwd17}
M(\lambda, \eta)= \pmatrix{e^{\eta/2} & 0  \cr 0 & e^{-\eta/2}}
\pmatrix{\cosh(\lambda/2) & \sinh(\lambda/2)  \cr \sinh(\lambda/2) &
  \cos(\lambda/2)}
\pmatrix{e^{-\eta/2} & 0  \cr 0 & e^{\eta/2}},
\end{equation}
respectively.  In view of the physical examples given in Sec.~\ref{isomor},
we shall call this the ``Wigner decomposition.''
Unlike the Bargmann decomposition, the Wigner decomposition is in the
form of a similarity transformation.

We note that both Eq.(\ref{bwd16}) and Eq.(\ref{bwd17}) are
written as similarity transformations.  Thus
\begin{eqnarray}\label{mass44}
&{}& [M(\theta, \eta)]^n
    = \pmatrix{\cos(n\theta/2) & - e^{-\eta}\sin(n\theta/2) \cr
        e^{\eta} \sin(n\theta/2) & \cos(n\theta/2)} ,  \nonumber \\[2ex]
&{}& [M(\lambda, \eta)]^n
    = \pmatrix{\cosh(n\lambda/2) & e^{\eta}\sinh(n\lambda/2)  \cr
    e^{-\eta} \sinh(n\lambda/2) & \cosh(n\lambda/2)}     , \nonumber \\[2ex]
&{}& [M(\gamma)]^n = \pmatrix{ 1 & 0 \cr n\gamma & 1} .
\end{eqnarray}
These expressions are useful for studying periodic
systems~\cite{bk13mop}.

\par

The question is what physics these decompositions describe in the
real world.  To address this, we study what the
Lorentz group does in the real world, and study isomorphism between the
$Sp(2)$ group and the Lorentz group applicable to the three-dimensional
space consisting of one time and two space coordinates.

\subsection{Isomorphism with the Lorentz group}\label{isomor}

The purpose of this section is to give physical interpretations of
the mathematical formulas given in Subsec.~\ref{bwdecom}.  We will
interpret these formulae in terms of the Lorentz transformations
which are normally described by four-by-four matrices.  For this
purpose, it is necessary to establish a correspondence between the
two-by-two representation of Sec.~\ref{bwdecom} and the four-by-four
representations of the Lorentz group.

\par

Let us consider the Minkowskian space-time four-vector
\begin{equation}
(t, z, x, y)
\end{equation}
where $\left(t^2 - z^2 - x^2 - y^2\right)$ remains invariant under Lorentz
transformations.   The Lorentz group consists of four-by-four matrices
performing Lorentz transformations in the Minkowski space.
 \par
In order to give physical interpretations to the three two-by-two matrices given
in Eq.(\ref{roty22}), Eq.(\ref{boostx22}), and Eq.(\ref{boostz22}),  we consider
rotations around the $y$ axis, boosts along the $x$ axis, and boosts along the $z$
axis.  The transformation is restricted in the three-dimensional subspace of
$(t, z, x)$.   It is then straight-forward to construct those four-by-four
transformation matrices where the $y$ coordinate remains invariant.  They are
given in Table~\ref{tab12}.  Their generators also given.  Those four-by-four
generators satisfy the Lie algebra given in Eq.(~\ref{sp2lie11}).

%--------------------------------------------------------------------------

\begin{center}
\begin{table}[thb]
\caption{Matrices in the two-by-two representation, and their corresponding
four-by-four generators and transformation matrices.}\label{tab12}
\vspace{2mm}
\begin{center}
\begin{tabular}{lllclccc}
\hline
\hline\\[-0.1ex]
 \hspace{1mm}& Matrices &\hspace{2mm} & Generators &\hspace{2mm}& Four-by-four & \hspace{2mm}& Transform Matrices{}\\[0.8ex]
\hline\\ [-0.8ex]
\hspace{1mm}& $ R(\theta)$ &\hspace{2mm} &  $J_{2} = \frac{1}{2}\pmatrix{0 & -i \cr i & 0}$ & \hspace{2mm}
  & $ \pmatrix{0 & 0 & 0 & 0 \cr 0 & 0 & -i & 0 \cr 0 & i & 0 & 0
      \cr  0 & 0 & 0 & 0 } $ &\hspace{2mm}&
$\pmatrix{1 & 0 & 0 & 0 \cr 0 & \cos\theta & -\sin\theta & 0 \cr
  0 & \sin\theta & \cos\theta & 0 \cr 0 & 0 & 0 & 1} $ \\[5.0ex]
\hline\\ [-0.8ex]
\hspace{1mm}& $ B(\eta)   $ &\hspace{2mm} &  $ K_{3} = \frac{1}{2}\pmatrix{i & 0 \cr 0 & -i}$ & \hspace{2mm}
  & $ \pmatrix{0 & i & 0 & 0 \cr i & 0 & 0 & 0 \cr 0 & 0 & 0 & 0 \cr  0 & 0 & 0 & 0} $ &\hspace{2mm}&
$ \pmatrix{\cosh\eta & \sinh\eta  & 0 & 0 \cr \sinh\eta & \cosh\eta & 0 & 0
\cr 0 & 0 & 1 & 0 \cr 0 & 0 & 0 & 1}$ \\[5.0ex]
\hline\\ [-0.8ex]
\hspace{1mm}& $ S(\lambda) $ &\hspace{2mm} &  $ K_1 = \frac{1}{2}\pmatrix{0 & i \cr i & 0}$ & \hspace{2mm}
  & $ \pmatrix{0 & 0 & i & 0 \cr 0 & 0 & 0 & 0 \cr i & 0 & 0 & 0 \cr  0 & 0 & 0 & 0} $ & \hspace{2mm}&
$ \pmatrix{\cosh\lambda & 0 & \sinh\lambda  & 0 \cr 0 & 1 & 0 & 0 \cr
 \sinh\lambda & 0 & \cosh\lambda & 0 \cr 0 & 0 & 0 & 1}$
  \\[5.0ex]
\hline
\hline
\end{tabular}

\end{center}
\end{table}
\end{center}
%----------------------------------------------------------------------------------

\section{Internal Space-time Symmetries}\label{internal}

We have seen that there corresponds a two-by-two matrix for each
four-by-four Lorentz transformation matrix.  It is possible to give
physical interpretations to those four-by-four matrices.  It must
thus be possible to attach a physical interpretation to each two-by-two
matrix.
\par
Since 1939~\cite{wig39} when Wigner introduced the concept of the little groups
many papers have been published  on
this subject, but most of them were based on the four-by-four
representation.
In this section, we shall give the formalism of little groups
in the language of two-by-two matrices.  In so doing, we
provide physical interpretations to the Bargmann and
Wigner decompositions introduced in Sec.~\ref{bwdecom}.

\par

\subsection{Wigner's Little Groups}\label{wiglittle}

In~\cite{wig39}, Wigner started with a free relativistic
particle with momentum, then constructed subgroups of the Lorentz
group whose transformations leave the four-momentum invariant.  These
subgroups thus define the internal space-time symmetry of the given
particle.  Without loss of generality, we assume that the particle
momentum is along the $z$ direction.  Thus rotations around
the momentum leave the momentum invariant, and this degree of freedom
defines the helicity, or the spin parallel to the momentum.

We shall use the word ``Wigner transformation'' for the transformation which
leaves the four-momentum invariant
\par
\begin{itemize}
\item[1.] For a massive particle, it is possible to find a Lorentz frame
where it is at rest with zero momentum.  The four-momentum can be written as
$m(1, 0, 0, 0)$,
where $m$ is the mass.  This four-momentum is
invariant under rotations in the three-dimensional $(z, x, y)$ space.
\par
\item[2.] For an imaginary-mass particle, there is the Lorentz frame where the
energy component vanishes.  The momentum four-vector can be written as
$p(0, 1, 0, 0)$ ,
where $p$ is the magnitude of the momentum.
\par
\item[3.] If the particle is massless, its four-momentum becomes
$ p(1, 1, 0, 0)$.
Here the first  and second components are equal in magnitude.
\end{itemize}
\noindent The constant factors in these four-momenta
do not play any significant roles.
Thus we write them as $(1, 0, 0, 0), (0, 1, 0, 0) $, and
$(1, 1, 0, 0)$ respectively.  Since Wigner worked with these three specific
four-momenta~\cite{wig39}, we call them Wigner four-vectors.

\par

All of these four-vectors are invariant under rotations around
the $z$ axis.  The rotation matrix is
\begin{equation}\label{rotz44}
Z(\phi) = \pmatrix{1  &  0 & 0 & 0 \cr
        0 & 1 & 0 & 0 \cr
     0 & 0 & \cos\phi  & -\sin\phi \cr 0 & 0 & \sin\phi & \cos\phi}     .
\end{equation}
In addition, the four-momentum of a massive particle is invariant under
the rotation around the $y$ axis, whose four-by-four matrix was given in
Table~\ref{tab12}.  The four-momentum of an imaginary particle is invariant under
the boost matrix $S(\lambda)$ given in Table~\ref{tab12}.   The problem
for the massless particle is more complicated, but will be discussed in
detail in Sec.~\ref{massless}. See Table~\ref{tab15}.
\par

%--------------------------------------------------------------------------
\begin{center}
\begin{table}[ht]
\caption{Wigner four-vectors and Wigner transformation matrices applicable
to two space-like and one time-like dimensions.  Each Wigner four-vector
remains invariant under the application of its Wigner matrix.
}\label{tab15}
\vspace{2mm}
\begin{center}
\begin{tabular}{lccccc}
\hline
\hline\\[-0.8ex]
 \hspace{3mm}& Mass & \hspace{3mm} & Wigner Four-vector &\hspace{3mm}
 & Wigner Transformation \\[1ex]
\hline\\[-0.8ex]
\hspace{3mm} & Massive  & \hspace{3mm} &
$(1, 0, 0, 0)$ & \hspace{3mm}
  & $\pmatrix{1 & 0 & 0 & 0 \cr 0 & \cos\theta & -\sin\theta & 0 \cr
  0 & \sin\theta & \cos\theta & 0 \cr  0 & 0 & 0 & 1} $
  \\[5.0ex]
\hline\\ [-0.8ex]
\hspace{3mm}& Massless & \hspace{3mm} &
 $ (1,1,0,0)$ &  \hspace{3mm}
  & $ \pmatrix{1 + \gamma^2/2  & -\gamma^2/2  & \gamma & 0 \cr
 \gamma^2/2 & 1 - \gamma^2/2 & \gamma & 0 \cr
 -\gamma & \gamma & 1 & 0 \cr  0 & 0 & 0 & 1} $ \\[5.0ex]
\hline\\ [-0.8ex]
\hspace{3mm}&  Imaginary mass  &\hspace{3mm}
&  $(0, 1, 0, 0)$  & \hspace{3mm}
  & $\pmatrix{\cosh\lambda  & 0 &\sinh\lambda & 0 \cr 0 & 1 & 0 & 0 \cr
  \sinh\lambda & 0 &\cosh\lambda & 0 \cr  0 & 0 & 0 & 1} $
  \\[5.0ex]
\hline
\hline
\end{tabular}

\end{center}
\end{table}
\end{center}
%----------------------------------------------------------------------------------

\subsection{Two-by-two Formulation of Lorentz Transformations}\label{2by2}
The Lorentz group is a group of four-by-four matrices performing Lorentz
transformations on the Minkowskian vector space of $(t, z, x, y),$ leaving
the quantity
\begin{equation}\label{4vec02}
t^2 - z^2 - x^2 - y^2
\end{equation}
invariant.  It is possible to perform the same transformation using
two-by-two matrices~\cite{knp86,kns13,naimark64}.

\par
In this two-by-two representation, the four-vector is written as
\begin{equation}\label{2b2}
X = \pmatrix{t + z  &  x - iy \cr x + iy & t - z} ,
\end{equation}
where its determinant is precisely the quantity given in Eq.(\ref{4vec02})
and the Lorentz transformation on this matrix is a determinant-preserving,
or unimodular transformation.  Let us consider the transformation matrix
as~\cite{kns13,naimark64}
\begin{equation}\label{alphabeta}
 G = \pmatrix{\alpha & \beta \cr \gamma & \delta}, \quad\mbox{and}\quad G^{\dagger} =
  \pmatrix{\alpha^* & \gamma^* \cr \beta^* & \delta^*} ,
\end{equation}
with
\begin{equation}\label{detone}
    \det{(G)} = 1 ,
\end{equation}
and the transformation
\begin{equation}\label{naim}
X' = G X G^{\dagger} .
\end{equation}
Since $G$ is not a unitary matrix, Eq.(\ref{naim}) not a unitary transformation,
but rather we call this the ``Hermitian
transformation''.   Eq.(\ref{naim}) can be written as
\begin{equation}\label{lt01}
\pmatrix{t' + z' & x' - iy' \cr x + iy & t' - z'}
 = \pmatrix{\alpha & \beta \cr \gamma & \delta}
  \pmatrix{t + z & x - iy \cr x + iy & t - z}
  \pmatrix{\alpha^* & \gamma^* \cr \beta^* & \delta^*} ,
\end{equation}
It is still a determinant-preserving unimodular transformation,
thus it is possible to write this as a  four-by-four transformation
matrix applicable to the four-vector $(t, z, x, y)$~\cite{knp86,kns13}.
\par

Since the $G$ matrix starts with four complex numbers and its determinant
is one by Eq.(\ref{detone}), it has six independent parameters.  The
group of these $G$ matrices is known to be locally isomorphic to the
group of four-by-four matrices performing Lorentz transformations on
the four-vector $(t, z, x, y)$.  In other words, for each $G$ matrix
there is a corresponding four-by-four Lorentz-transform matrix~\cite{kns13}.
\par

The matrix $G$ is not a unitary matrix, because its Hermitian conjugate
is not always its inverse.  This group has a unitary
subgroup called $SU(2)$ and another consisting only of real matrices
called $Sp(2)$.  For this later subgroup, it is sufficient
to work with the three matrices $R(\theta), S(\lambda)$, and $B(\eta)$ given
in Eqs.(\ref{roty22}), (\ref{boostx22}), and (\ref{boostz22}) respectively.
Each of these matrices has its corresponding four-by-four matrix applicable
to the $(t, z, x, y)$.   These matrices with their four-by-four counterparts are
tabulated in Table~\ref{tab12}.

\par

The energy-momentum four vector can also be written as a two-by-two matrix.
It can be written as
\begin{equation}\label{mom00}
P = \pmatrix{p_0 + p_z & p_x - ip_y \cr p_x + ip_y & p_0 - p_z} ,
\end{equation}
with
\begin{equation}
\det{(P)} = p_0^2 - p_x^2 - p_y^2 - p_z^2,
\end{equation}
which means
\begin{equation}\label{mass}
\det{(P)} = m^2,
\end{equation}
where $m$ is the particle mass.
\par
The Lorentz transformation can be written explicitly as
 \begin{equation}
 P' = G P G^{\dagger} ,
 \end{equation}
or
\begin{equation}\label{lt03}
\pmatrix{p_0' + p_z' & p_x' - ip_y' \cr p'_x + ip'_y & E' - p'_z}
 = \pmatrix{\alpha & \beta \cr \gamma & \delta}
  \pmatrix{p_0 + p_z & p_x - ip_y \cr p_x + ip_y & p_0 - p_z}
  \pmatrix{\alpha^* & \gamma^* \cr \beta^* & \delta^*} .
\end{equation}
This is an unimodular transformation, and the mass is a Lorentz-invariant
variable.  Furthermore, it was shown in~\cite{kns13} that Wigner's little
groups for massive, massless, and imaginary-mass particles can be explicitly
defined in terms of two-by-two matrices.
\par
Wigner's little group consists of two-by-two
matrices satisfying
\begin{equation}\label{wigcondi}
P = W P W^{\dagger} .
\end{equation}
The two-by-two $W$ matrix is not an identity matrix, but tells
about the internal space-time symmetry of a particle with a given
energy-momentum four-vector.  This aspect was not known when Einstein
formulated his special relativity in 1905, hence  the internal space-time
symmetry was not an issue at that time.  We  call the two-by-two
matrix $W$ the Wigner matrix, and call the condition of Eq.(\ref{wigcondi})
the Wigner condition.

\par

If determinant of $W$ is a positive number, then $P$ is proportional to
\begin{equation}\label{massive}
         P = \pmatrix{1 & 0 \cr 0 & 1},
\end{equation}
corresponding to a massive particle at rest, while if
the determinant is negative, it is
proportional to
\begin{equation}\label{superlum}
         P = \pmatrix{1 & 0 \cr 0 & -1} ,
\end{equation}
corresponding to an imaginary-mass particle moving faster than light
along the $z$ direction, with a vanishing energy component.
If the determinant is zero,  $P$ is
\begin{equation}\label{mzero}
         P = \pmatrix{1 & 0 \cr 0 & 0} ,
\end{equation}
which is proportional to the four-momentum matrix for a massless particle
moving along the $z$ direction.
\par
For all three cases, the matrix of the form
\begin{equation}\label{rotz22}
Z(\phi) = \pmatrix{e^{-i\phi/2} & 0 \cr 0 & e^{i\phi/2}}
\end{equation}
will satisfy the Wigner condition of Eq.(\ref{wigcondi}).  This matrix
corresponds to rotations around the $z$ axis.

\par
For the massive particle with the four-momentum of Eq.(\ref{massive}),
the transformations with the rotation matrix of Eq.(\ref{roty22})
leave the $P$ matrix of Eq.(\ref{massive}) invariant.  Together
with the $Z(\phi)$ matrix, this rotation matrix leads to the subgroup
consisting of the unitary subset of the $G$ matrices.  The unitary subset
of $G$ is $SU(2)$ corresponding to the three-dimensional rotation group
dictating the spin of the particle~\cite{knp86}.
\par
For the massless case, the transformations with the triangular matrix
of the form
\begin{equation}\label{trian}
\pmatrix{1 & \gamma \cr 0 & 1}
\end{equation}
leave the momentum matrix of Eq.(\ref{mzero}) invariant.  The physics
of this matrix has a stormy history, and the variable $\gamma$
leads to a gauge transformation applicable to massless
particles~\cite{hks82,hks86ajp,kiwi87jmp,kiwi90jmp}.
\par
For a particle with an imaginary mass, a $W$ matrix of the form of Eq.(\ref{boostx22})
leaves the four-momentum of Eq.(\ref{superlum}) invariant.
\par
Table~\ref{tab11} summarizes the transformation matrices for
Wigner's little groups for massive, massless, and imaginary-mass particles.
Furthermore, in terms of their traces, the matrices given in this
subsection can be compared with those given in Subsec.~\ref{mform} for
the damped oscillator.  The comparisons are given in Table~\ref{tab14}.
\par

Of course, it is a challenging problem to have one expression for all
three classes. This problem has been discussed in the
literature~\cite{bk10jmo}, and the damped oscillator case of
Sec.~\ref{dampo} addresses the continuity problem.

%--------------------------------------------------------------------------

\begin{center}
\begin{table}[ht]
\caption{Wigner vectors and Wigner matrices in the two-by-two representation.
  The trace of the matrix tells whether the particle $m^2$ is positive,
  zero, or negative. }\label{tab11}
\vspace{2mm}
\begin{center}
\begin{tabular}{llclccl}
\hline
\hline \\[0.5ex]
 Particle mass &{}&  Four-momentum  & {} &  Transform matrix &{}& Trace
 \\[1.0ex]
\hline\\
Massive  &{}& $\pmatrix{1 & 0 \cr 0 & 1}$
&{}&
$\pmatrix{\cos(\theta/2) & -\sin(\theta/2)\cr \sin(\theta/2) & \cos(\theta/2)}$
&{}&  less than 2
\\[4ex]
Massless  &{}&
$\pmatrix{1 & 0 \cr 0 & 0}$
&{}& $\pmatrix{1 & \gamma \cr 0 & 1}$
&{}&  equal to 2
\\[4ex]
Imaginary mass &{}&
$\pmatrix{1 & 0\cr 0 & -1}$
&{}&  $\pmatrix{\cosh(\lambda/2) & \sinh(\lambda/2) \cr \sinh(\lambda/2) & \cosh(\lambda/2)}$
&{}& greater than 2
\\[4ex]
\hline
\hline\\[-0.8ex]
\end{tabular}
\end{center}
\end{table}
\end{center}
%-------------------------------------------------------------------------------------------

%----------------------------------------------------------------------------------
\begin{center}
\begin{table}[ht]
\caption{Damped Oscillators and Space-time Symmetries.  Both share $Sp(2)$
as their symmetry group.}\label{tab14}
\vspace{2mm}
\begin{center}
\begin{tabular}{llllll}
\hline\\ [-2.5ex]
\hline\\ [-0.5ex]
 \hspace{3mm}& Trace &\hspace{5mm} & Damped Oscillator &\hspace{3mm}  & Particle Symmetry \\
\hline\\
\hspace{3mm}& Smaller than 2 &\hspace{5mm}&  Oscillation Mode &\hspace{3mm}  &  Massive Particles
\\[2ex]
\hline \\[1ex]
\hspace{3mm} & Equal to 2 &\hspace{5mm} & Transition Mode &\hspace{3mm}  &  Massless Particles
\\[2ex]
\hline \\[1ex]
\hspace{3mm} & Larger than 2 &\hspace{5mm} & Damping Mode &\hspace{3mm}    &   Imaginary-mass Particles
\\[2ex]
\hline
\hline
\end{tabular}
\end{center}
\end{table}
\end{center}
%----------------------------------------------------------------------------------

\section{Lorentz Completion of Wigner's Little Groups}\label{complete}

So far we have considered transformations applicable only to
$(t, z, x)$ space.  In order to study the full symmetry, we have to
consider rotations around the $z$ axis.  As previously stated, when a
particle moves along
this axis, this rotation defines the helicity of the particle.
\par
In~\cite{wig39}, Wigner worked out the little group
of a massive particle at rest.  When the particle
gains a momentum along the $z$ direction,
the single particle can reverse the direction of momentum, the spin, or
both.  What happens to the internal space-time symmetries
is discussed in this section.

\subsection{Rotation around the $z$ axis}\label{rotz}

In Sec.~\ref{sp2}, our kinematics was restricted to the two-dimensional
space of $z$ and $x$, and thus includes rotations around the $y$ axis.
We now introduce the four-by-four matrix of Eq.(\ref{rotz44})
performing rotations around the $z$ axis.  Its corresponding two-by-two
matrix was given in Eq.(\ref{rotz22}).
Its generator is
\begin{equation}
J_3 = \frac{1}{2}\pmatrix{1 & 0 \cr 0 & -1}.
\end{equation}
If we introduce this  additional matrix for the three generators we used
in Secs.~\ref{sp2} and \ref{bwdecom}, we end up the closed set of commutation relations
\begin{equation}\label{sl2clie}
[J_{i},J_{j}]=i \epsilon_{ijk} J_{k},\qquad
[J_{i},K_{j}]=i \epsilon_{ijk}K_{k},\qquad
[K_{i},K_{j}]= -i \epsilon_{ijk} J_{k},
\end{equation}
with
\begin{equation}
J_{i} = \frac{1}{2}\sigma_{i}, \quad\mbox{and}\quad K_{i} = \frac{i}{2}\sigma_{i} ,
\end{equation}
where $\sigma_{i}$ are the two-by-two Pauli spin matrices.
\par
For each of these two-by-two matrices there is a corresponding four-by-four
matrix generating Lorentz transformations on the four-dimensional Lorentz
group.  When these two-by-two matrices are imaginary, the corresponding
four-by-four matrices were given in Table~\ref{tab12}.  If they are real,
the corresponding
four-by-four matrices were given in Table~\ref{tab16}.

%--------------------------------------------------------------------------
\begin{center}
\begin{table}[ht]
\caption{Two-by-two and four-by-four generators not included in
Table~\ref{tab12}.  The generators given there and given here constitute
the set of six generators for $SL(2,c)$ or of the Lorentz group given
in Eq.(~\ref{sl2clie}).}\label{tab16}
\vspace{2mm}
\begin{center}
\begin{tabular}{lllclc}
\hline
\hline\\[-0.1ex]
 \hspace{3mm}& Generator &\hspace{3mm} & Two-by-two&\hspace{3mm}  & Four-by-four \\[0.8ex]
\hline\\ [-0.8ex]
\hspace{3mm}& $ J_3$ &\hspace{3mm} &  $\frac{1}{2}\pmatrix{1 & 0 \cr 0 & -1}$ & \hspace{3mm}
  & $\pmatrix{0 & 0 & 0 & 0 \cr 0 & 0 & 0 & 0 \cr 0 & 0 & 0 & -i \cr  0 & 0 & i & 0} $
  \\[5.0ex]
\hline\\ [-0.8ex]
\hspace{3mm}& $ J_1   $ &\hspace{3mm} &  $\frac{1}{2}\pmatrix{0 & 1 \cr 1 & 0}$ & \hspace{3mm}
  & $\pmatrix{0 & 0 & 0 & 0 \cr 0 & 0 & 0 & i \cr 0 & 0 & 0 & 0 \cr  0 & -i & 0 & 0} $
  \\[5.0ex]
\hline\\ [-0.8ex]
\hspace{3mm}& $ K_2 $ &\hspace{3mm} &  $\frac{1}{2}\pmatrix{0 & 1 \cr -1 & 0}$ & \hspace{3mm}
  & $\pmatrix{0 & 0 & 0 & i \cr 0 & 0 & 0 & 0 \cr 0 & 0 & 0 & 0 \cr  i & 0 & 0 & 0} $
  \\[5.0ex]
\hline
\hline
\end{tabular}

\end{center}
\end{table}
\end{center}
%----------------------------------------------------------------------------------

This set of commutation relations is known as the Lie algebra for
the SL(2,c), namely the group of two-by-two elements with unit
determinants.  Their elements are complex.   This set is also
the Lorentz group performing Lorentz transformations on the
four-dimensional Minkowski space.
\par
This set has many useful subgroups.  For the group $SL(2,c)$, there is
a subgroup consisting only of real matrices, generated by
the two-by-two matrices given in Table~\ref{tab12}.  This three-parameter
subgroup is precisely the $Sp(2)$ group we used in
Secs.~\ref{sp2} and \ref{bwdecom}.  Their generators satisfy the
Lie algebra given in Eq.(\ref{sp2lie11}).

\par

In addition, this group has the following Wigner subgroups governing
the internal space-time symmetries of particles in the Lorentz-covariant
world~\cite{wig39}:

\par

\par
\begin{itemize}

\item[1.] The $J_i$ matrices form a closed set of commutation relations.
The subgroup generated by these Hermitian matrices is $SU(2)$ for electron
spins. The corresponding rotation group
does not change the four-momentum of the particle at rest.
This is Wigner's little group for massive particles.
\par

If the particle is at rest, the two-by-two form of
the four-vector is given by Eq.(\ref{massive}).
The Lorentz transformation generated by $J_3$ takes the form
\begin{equation}
\pmatrix{e^{i\phi/2} & 0 \cr 0 & e^{-i\phi/2}}
\pmatrix{1 & 0 \cr 0 & 1}
\pmatrix{e^{-i\phi/2} & 0 \cr 0 & e^{i\phi/2}}
= \pmatrix{1 & 0 \cr 0 & 1}
\end{equation}
Similar computations can be carried out for $J_1$ and $J_2$.

\par
\item[2.] There is another $Sp(2)$ subgroup, generated by $K_1, K_2$, and
$J_3$. They satisfy the commutation relations
\begin{equation}\label{sp2lie22}
 \left[K_1, K_2\right] = -i J_3, \qquad \left[J_3, K_1\right] = i K_2, \qquad
 \left[K_2, J_3\right] = i K_1 .
\end{equation}
The Wigner transformation generated by these two-by-two matrices
leave the momentum four-vector of Eq.(\ref{superlum})
invariant.  For instance, the transformation matrix generated by
$K_2$ takes the form
\begin{equation}
\exp{\left( -i\xi K_2 \right)} =
\pmatrix{\cosh(\xi/2) & i\sinh(\xi/2) \cr
  i\sinh(\xi/2) & \cosh(\xi/2) }
\end{equation}
and the Wigner transformation takes the form
\begin{equation}
\pmatrix{\cosh(\xi/2) & i\sinh(\xi/2) \cr -i\sinh(\xi/2) & \cosh(\xi/2) }
\pmatrix{1 & 0 \cr 0 & -1}
\pmatrix{\cosh(\xi/2) & i\sinh(\xi/2) \cr -i\sinh(\xi/2) & \cosh(\xi/2)}
= \pmatrix{1 & 0 \cr 0 & -1} .
\end{equation}
Computations with $K_2$ and $J_3$ lead to the same result.
\par
Since the determinant of the four-momentum matrix is negative,
the particle has an imaginary mass.  In the language of the
four-by-four matrix, the transformation matrices leave the four-momentum
of the form $(0, 1, 0, 0)$ invariant.

\item[3.] Furthermore, we can consider the following combinations
of the generators:
\begin{equation}\label{e211}
  N_1  = K_1 - J_2 = \pmatrix{0 &  i \cr 0 & 0}, \quad\mbox{and}\quad
 N_2 = K_2 + J_1 = \pmatrix{0 & 1 \cr 0 & 0} .
\end{equation}
Together with $J_3$, they satisfy the the following commutation relations.
\begin{equation}\label{e212}
\left[N_1, N_2\right] = 0, \qquad \left[N_1, J_3\right] = -i N_2,
\qquad \left[N_2, J_3\right] = i N_1,
\end{equation}
In order to understand this set of commutation relations, we can consider
an $x~y$ coordinate system in a two-dimensional space. Then
rotation around the origin is generated by
\begin{equation}\label{e213}
J_3 = -i \left(x \frac{\partial}{\partial y} -
y \frac{\partial}{\partial x} \right) ,
\end{equation}
and the two translations are generated by
\begin{equation}\label{e215}
N_1 = -i \frac{\partial}{\partial x}, \quad\mbox{and}\quad
N_2 = -i \frac{\partial}{\partial y},
\end{equation}
for the $x$ and $y$ directions respectively.  These operators
satisfy the commutations relations given in Eq.(\ref{e212}).
\end{itemize}
\par

The two-by-two matrices of Eq.(\ref{e211}) generate the following
transformation matrix.
\begin{equation}\label{e216}
G(\gamma, \phi) =
\exp{\left[-i\gamma \left( N_1 \cos\phi + N_2\sin\phi\right)\right]} =
\pmatrix{1 & \gamma e^{-i\phi} \cr  0 & 1}.
\end{equation}
The two-by-two form for the four-momentum for the massless particle is
given by Eq.(\ref{mzero}).
The computation of the Hermitian transformation using
this matrix is
\begin{equation}
\pmatrix{1 & \gamma e^{-i\phi} \cr  0 & 1}
\pmatrix{1 & 0 \cr 0 & 0}
\pmatrix{1 & 0 \cr \gamma e^{i\phi}   & 1}
  = \pmatrix{1 & 0 \cr 0 & 0} ,
\end{equation}
confirming that $N_1$ and $N_2$, together with $J_{3}$, are
the generators of the $E(2)$-like little group for massless particles
in the two-by-two representation.  The transformation that does this
in the physical world is described in the
following section.

\subsection{E(2)-like Symmetry of Massless Particles}\label{e2symm}

From the four-by-four generators of $K_{1,2}$ and $J_{1,2},$ we can write
\begin{equation}
  N_1 = \pmatrix{0 & 0 & i & 0 \cr 0 & 0 & i & 0 \cr
           i & -i & 0 & 0 \cr 0 &  0 & 0 & 0 } ,
           \quad\mbox{and}\quad
  N_2 = \pmatrix{0 & 0 & 0 & i \cr 0 & 0 & 0 & i \cr
         0 & 0 & 0 & 0 \cr i &  -i & 0 & 0 }   .
\end{equation}
These matrices lead  to the transformation matrix of the form
\begin{equation}\label{e231}
G(\gamma, \phi) = \pmatrix{1 + \gamma^2/2  & - \gamma^2/2   &
  \gamma \cos\phi & \gamma \sin\phi \cr
 \gamma^2/2  &  1 - \gamma^2/2  & \gamma\cos\phi & \gamma\sin\phi \cr
 -\gamma \cos\phi & \gamma \cos\phi & 1 & 0 \cr
 -\gamma \sin\phi &  \gamma \sin\phi  & 0 & 1}
\end{equation}
This matrix leaves the four-momentum invariant, as we can
see from
\begin{equation}
G(\gamma, \phi)\pmatrix{1 \cr 1 \cr 0 \cr 0}
 = \pmatrix{1 \cr 1 \cr 0 \cr 0} .
\end{equation}
When it is applied to the photon four-potential
\begin{equation}\label{e235}
G(\gamma, \phi)
\pmatrix{A_0 \cr A_3 \cr A_1 \cr A_2} =
\pmatrix{A_0 \cr A_3 \cr A_1 \cr A_2} +
\gamma\left(A_1\cos\phi + A_2\sin\phi\right)\pmatrix{1 \cr 1 \cr 0 \cr 0} ,
\end{equation}
with the Lorentz condition which leads to $A_{3} = A_{0}$ in the zero
mass case.
Gauge transformations are well known for electromagnetic
fields and photons.  Thus Wigner's
little group leads to gauge transformations.
\par
In the two-by-two representation, the electromagnetic four-potential
takes the form
\begin{equation}\label{e237}
\pmatrix{2A_0 & A_1 - iA_2 \cr A_1 + iA_2 & 0},
\end{equation}
with the Lorentz condition $A_3 = A_0$.
Then the two-by-two form of  Eq.(\ref{e235}) is
\begin{equation}\label{e239}
\pmatrix{1 & \gamma e^{-i\phi} \cr 0 & 1}
\pmatrix{2A_0 & A_1 - iA_2 \cr A_1 + iA_2 & 0}
\pmatrix{1 & 0 \cr \gamma e^{i\phi}  & 1} ,
\end{equation}
which becomes
\begin{equation}\label{e241}
\pmatrix{A_0 & A_1 - iA_2 \cr A_1 + iA_2 & 0}
+ \pmatrix{2\gamma\left(A_1 \cos\phi - A_2 \sin\phi\right) & 0 \cr 0 & 0} .
\end{equation}
This is the two-by-two equivalent of the gauge transformation given
in Eq.(\ref{e235}).

\par
For massless spin-1/2 particles starting with the two-by-two expression of
$G(\gamma, \phi)$ given in Eq.(\ref{e216}), and
considering the spinors
\begin{equation}\label{e242}
  u = \pmatrix{1 \cr 0}, \quad\mbox{and}\quad v = \pmatrix{0 \cr 1} ,
\end{equation}
for spin-up and spin-down states respectively,
\begin{equation} \label{e244}
G u = u, \quad\mbox{and}\quad
    G v = v + \gamma e^{-i\phi} u,
\end{equation}
This means that the spinor $u$ for spin up is invariant under the gauge
transformation while $v$ is not.  Thus, the polarization of massless
spin-1/2 particle, such as neutrinos, is a consequence of the gauge
invariance.  We shall continue this discussion in Sec.~\ref{massless}.

\subsection{Boosts along the $z$ axis}\label{boostz}

In Subsec.~\ref{wiglittle} and Subsec.~\ref{rotz}, we studied
Wigner transformations for fixed values of the four-momenta. The next
question is what happens when the system is boosted along the
$z$ direction, with the transformation
\begin{equation} \label{boost01}
\pmatrix{t' \cr z'} =
\pmatrix{\cosh\eta & \sinh\eta \cr \sinh\eta & \cosh\eta}
\pmatrix{t \cr z} .
\end{equation}
Then the four-momenta become
\begin{equation} \label{boost03}
(\cosh\eta, \sinh\eta, 0, 0), \quad
(\sinh\eta, \cosh\eta, 0, 0), \quad
  e^{\eta}(1, 1,  0, 0),
\end{equation}
respectively for massive, imaginary, and massless particles cases.
In the two-by-two representation, the boost matrix is
\begin{equation}\label{boost05}
\pmatrix{e^{\eta/2} & 0 \cr  0 & e^{-\eta/2}} ,
\end{equation}
and the four-momenta of Eqs.(\ref{boost03}) become
\begin{equation} \label{boost07}
\pmatrix{e^{\eta} & 0 \cr 0 & e^{-\eta}}, \qquad
\pmatrix{e^{\eta} & 0 \cr 0 & -e^{-\eta}}, \qquad
\pmatrix{e^{\eta} & 0 \cr 0 & 0},
\end{equation}
respectively.
These matrices become  Eqs.(\ref{massive}), (\ref{superlum}), and
(\ref{mzero}) respectively when $\eta = 0$.

\par

\begin{figure}[thb]
\centerline{\includegraphics[scale=0.6]{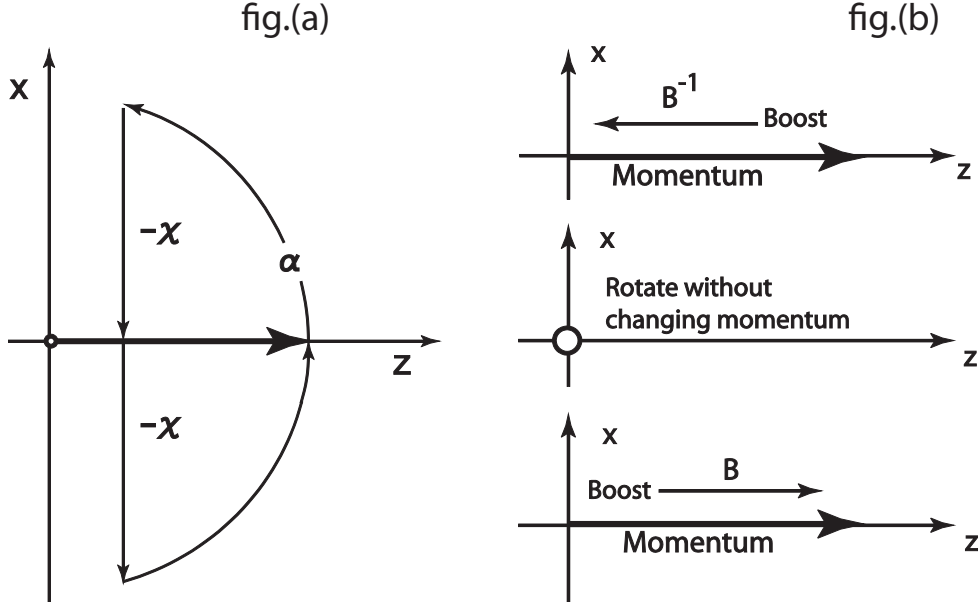}}
\caption{Bargmann and  Wigner Decompositions.   The Bargmann decomposition
is illustrated in fig.(a).  Starting from a momentum along the $z$ direction,
we can rotate, boost, and rotate to bring the momentum to the original
position.  The resulting matrix is the product of one boost and two
rotation matrices. In the Wigner decomposition, the particle is boosted
back to the frame where the Wigner transformation can be applied.  Make a
Wigner transformation there and come back to the original state of the
momentum as illustrated in fig.(b).  This process also can also be
written as the product of three simple matrices.}\label{bw11}
\end{figure}

We are interested in Lorentz transformations which leave a given non-zero
momentum invariant.  We can consider a Lorentz boost along the direction
preceded and followed by identical rotation matrices, as described in
Fig.(\ref{bw11}) and the transformation matrix as
\begin{equation}\label{boost09}
\pmatrix{\cos(\alpha/2) & -\sin(\alpha/2) \cr \sin(\alpha/2) & \cos(\alpha/2)}
\pmatrix{\cosh\chi & -\sinh\chi \cr -\sinh\chi & \cosh\chi}
\pmatrix{\cos(\alpha/2) & -\sin(\alpha/2) \cr \sin(\alpha/2) & \cos(\alpha/2)} ,
\end{equation}
which becomes
\begin{equation}\label{boost11}
\pmatrix{(\cos\alpha)\cosh\chi & -\sinh\chi - (\sin\alpha)\cosh\chi \cr
- \sinh\chi + (\sin\alpha)\cosh\chi & (\cos\alpha)\cosh\chi} .
\end{equation}

%--------------------------------------------------------------------
%----------------------------------------------------------------------

\par
Except the sign of $\chi$, the two-by-two matrices of Eq.(\ref{boost09})
and Eq.(\ref{boost11}) are identical with those given in Sec.~\ref{bwdecom}.
The only difference is the sign of the parameter $\chi$.  We are thus
ready to interpret this expression in terms of physics.

\par
\begin{itemize}
\item[1.] If the particle is massive,  the off-diagonal elements
of Eq.(\ref{boost11}) have opposite signs, and this matrix can be
decomposed into
\begin{equation} \label{boost21}
\pmatrix{e^{\eta/2} & 0 \cr  0 & e^{-\eta/2}}
  \pmatrix{\cos(\theta/2) & -\sin(\theta/2)  \cr
       \sin(\theta/2) & \cos(\theta/2)}
\pmatrix{e^{\eta/2} & 0 \cr  0 & e^{-\eta/2}} .
\end{equation}
with
\begin{equation} \label{bost21a}
\cos(\theta/2) = (\cosh\chi) \cos\alpha, \quad\mbox{and}\quad
e^{2\eta} = \frac{\cosh(\chi)\sin\alpha + \sinh\chi}
{\cosh(\chi)\sin\alpha - \sinh\chi} ,
\end{equation}
and
\begin{equation} \label{boost21b}
  e^{2\eta} = \frac{p_0 + p_z} {p_0 - p_z}.
\end{equation}

\par

According to Eq.(\ref{boost21}) the first matrix (far right) reduces
the particle momentum to zero.  The second matrix rotates the particle
without changing the momentum.  The third matrix boosts the particle
to restore its original momentum.  This is the extension of
Wigner's original idea to moving particles.

\par

\item[2.] If the particle has an imaginary mass, the off-diagonal elements
of Eq.(\ref{boost11}) have the same sign,
\begin{equation}\label{boost25}
\pmatrix{e^{\eta/2} & 0 \cr  0 & e^{-\eta/2}}
\pmatrix{\cosh(\lambda/2) & -\sinh(\lambda/2)  \cr
       \sinh(\lambda/2) & \cosh(\lambda/2)}
\pmatrix{e^{\eta/2} & 0 \cr  0 & e^{-\eta/2}} ,
\end{equation}
with
\begin{equation}\label{boost25a}
\cosh(\lambda/2) = (\cosh\chi) \cos\alpha, \quad\mbox{and}\quad
e^{2\eta} = \frac{\sinh\chi + \cosh(\chi)\sin\alpha}
{\cosh(\chi)\sin\alpha  - \sinh\chi},
\end{equation}
and
\begin{equation}\label{boost25b}
   e^{2\eta} = \frac{p_0 + p_z} {p_z - p_0}.
\end{equation}
This is also a three-step operation. The first matrix brings the
particle momentum to the zero-energy state with  $p_0 = 0$.
Boosts along the $x$ or $y$ direction do not change the four-momentum.
We can then boost the particle back to restore its momentum.   This
operation is also an extension of the Wigner's original little group.
Thus, it is quite appropriate to call the formulas of Eq.(\ref{boost21})
and Eq.(\ref{boost25}) Wigner decompositions.

\item[3.] If the particle mass is zero with
\begin{equation} \label{boost27a}
\sinh\chi = (\cosh\chi)\sin\alpha,
\end{equation}
the $\eta$ parameter becomes infinite, and the Wigner decomposition does not
appear to be useful.  We can then go back to the Bargmann decomposition of
Eq.(\ref{boost09}).  With the condition of Eq.(\ref{boost27a}),
Eq.(\ref{boost11}) becomes
\begin{equation}\label{boost27}
\pmatrix{1 & -\gamma \cr  0 & 1} ,
\end{equation}
with
\begin{equation} \label{boost27b}
   \gamma  = 2 \sinh\chi.
\end{equation}
The decomposition ending with a triangular matrix is called the Iwasawa
decomposition~\cite{iwa49,gk01} and its physical interpretation was given in
Subsec.~\ref{e2symm}.  The $\gamma$ parameter does not depend
on $\eta.$

\end{itemize}
\par
Thus, we have given physical interpretations to the Bargmann and Wigner
decompositions given in Sec.~(\ref{bwdecom}).
Consider what happens when the momentum becomes large.
Then $\eta$ becomes large for nonzero mass cases.  All three
four-momenta in Eq.(\ref{boost07}) become
\begin{equation}
   e^{\eta} \pmatrix{1 & 0 \cr 0 & 0} .
\end{equation}
As for the Bargmann-Wigner matrices, they become the triangular
matrix of Eq.(\ref{boost27}), with $\gamma = \sin(\theta/2) e^{\eta}$
and $\gamma = \sinh(\lambda/2) e^{\eta},$ respectively for the
massive and imaginary-mass cases.
\par

In Subsec.~\ref{e2symm}, we concluded that the triangular matrix
corresponds to gauge transformations. However, particles with imaginary mass
are not observed.  For massive particles, we can start with
the three-dimensional rotation group.  The rotation around the
$z$ axis is called helicity, and remains invariant under the
boost along the $z$ direction.  As for the transverse rotations,
they become gauge transformation as illustrated in Table~\ref{tab22}.

%--------------------------------------------------------------------
\begin{table}[thb]
\caption{Covariance of the energy-momentum relation, and covariance of
the internal space-time symmetry.  Under the Lorentz boost along the $z$
direction, $J_3$ remains invariant, and this invariant component of the
angular momentum is called the helicity.  The transverse component $J_1$
and $J_2$ collapse into a gauge transformation.  The $\gamma$ parameter
for the massless case has been studied in earlier papers in the
four-by-four matrix formulation~\cite{hks82,kiwi90jmp}.}\label{tab22}
\vspace{5mm}
\begin{center}
\begin{tabular}{ccccc}
\hline\hline \\[0.5ex]
Massive, Slow &\hspace{7mm} & COVARIANCE &\hspace{10mm}& Massless, Fast \\[2mm]
\hline\\
$E = p^{2}/2m$ &{}& Einstein's $E = mc^{2}$ &{}& $E = cp$ \\[4mm]
\hline \\
$J_{3}$ &{}& {}  &{}&   Helicity \\ [-1mm]
{} &{}& Wigner's Little Group &{}& {} \\[-1mm]
$J_{1}, J_{2}$ &{}& {} &{}& Gauge Transformation \\[4mm]
\hline\hline\\[-0.8ex]
\end{tabular}
\end{center}
\end{table}
%-----------------------------------------------------------------------

\subsection{Conjugate Transformations}\label{conju}

The most general form of the $SL(2,c)$ matrix is given in
Eq.(\ref{alphabeta}). Transformation operators for the Lorentz group
are given in exponential form as:
\begin{equation}\label{conju01}
D = \exp{\left\{ -i \sum_{i=1}^3 \left(\theta_i J_i + \eta_i K_i \right)\right\}},
\end{equation}
where the $J_i$ are the generators of rotations and the $K_i$ are the
generators of proper Lorentz boosts.  They satisfy the Lie algebra given
in Eq.(\ref{sp2lie11}).  This set of commutation relations is invariant
under the sign change of the boost generators $K_{i}$.  Thus, we can
consider ``dot conjugation'' defined as
\begin{equation}\label{conju03}
\dot{D} = \exp{\left\{ -i \sum_{i=1}^3
\left(\theta_i J_i - \eta_i K_i \right)\right\}} ,
\end{equation}
Since $K_{i}$ are anti-Hermitian while $J_{i}$ are Hermitian, the
Hermitian conjugate of the above expression is
\begin{equation}\label{conju05}
D^{\dagger} = \exp{\left\{ -i \sum_{i=1}^3 \left(-\theta_i J_i + \eta_i K_i \right)\right\}},
\end{equation}
while the Hermitian conjugate of $G$ is
\begin{equation}\label{conju07}
\dot{D}^{\dagger} = \exp{\left\{ -i \sum_{i=1}^3 \left(-\theta_i J_i - \eta_i K_i \right) \right\}},
\end{equation}

%----------------------------------------------------------------------------
\begin{figure}[thb]
\centerline{\includegraphics[scale=0.7]{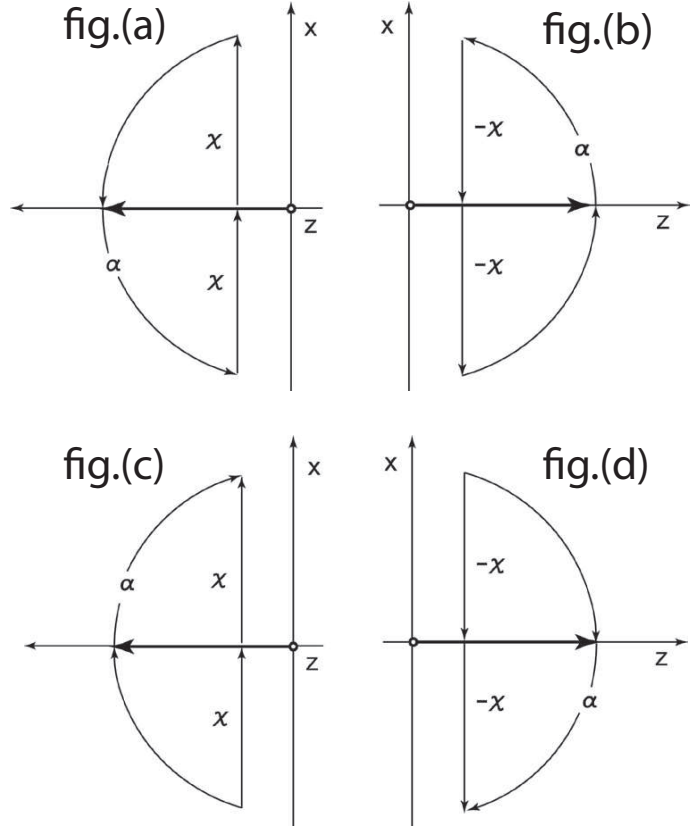}}
\caption{Four D-loops resulting from the Bargmann decomposition.  Let us
go back to Fig.~\ref{bw11}.  If we reverse of the direction of the boost,
the result is fig.(a).  From fig.(a), if we invert the space, we come back
to fig.(b).  If we reverse the direction of rotation from fig.(a), the
result is fig.(c).  If both the rotation and space are reversed, the result
is the fig.(d).}\label{dloops}
\end{figure}
%----------------------------------------------------------------------

Since we understand the rotation around the $z$ axis, we can now restrict
the kinematics to the $z t$ plane, and work with the $Sp(2)$ symmetry.
Then the $D$ matrices can be considered as Bargmann decompositions.
First, $D$ and $\dot{D}$, and their Hermitian conjugates are
\begin{eqnarray}\label{conju11}
&{}& D(\alpha, \chi) =
\pmatrix{(\cos\alpha)\cosh\chi &  \sinh\chi -(\sin\alpha)\cosh\chi \cr
  \sinh\chi + (\sin\alpha)\cosh\chi & (\cos\alpha)\cosh\chi} , \nonumber \\[2ex]
&{}&
\dot{D}(\alpha, \chi) =
\pmatrix{(\cos\alpha)\cosh\chi &  -\sinh\chi - (\sin\alpha)\cosh\chi \cr
 - \sinh\chi + (\sin\alpha)\cosh\chi & (\cos\alpha)\cosh\chi}.
\end{eqnarray}
These matrices correspond to the ``D loops'' given in fig.(a) and fig.(b)
of Fig.~\ref{dloops} respectively.  The ``dot'' conjugation changes
the direction of boosts. The dot conjugation leads to the inversion
of the space which is called the parity operation.
\par
We can also consider changing the direction of rotations.  Then they
result in the Hermitian conjugates.  We can write their matrices as
\begin{eqnarray}\label{conju12}
&{}&
D^{\dagger}(\alpha, \chi) =
\pmatrix{(\cos\alpha)\cosh\chi &  \sinh\chi + (\sin\alpha)\cosh\chi \cr
  \sinh\chi - (\sin\alpha)\cosh\chi & (\cos\alpha)\cosh\chi} ,  \nonumber \\[2ex]
&{}&
\dot{D}^{\dagger}(\alpha, \chi) =
\pmatrix{(\cos\alpha)\cosh\chi &  -\sinh\chi + (\sin\alpha)\cosh\chi \cr
 -\sinh\chi - (\sin\alpha)\cosh\chi & (\cos\alpha)\cosh\chi} .
\end{eqnarray}

From the exponential expressions from Eq.(\ref{conju01}) to Eq.(\ref{conju07}),
it is clear that
\begin{equation}
  D^{\dagger} = \dot{D}^{-1}, \quad\mbox{and}\quad
  \dot{D}^{\dagger} = D^{-1}.
\end{equation}
The D loop given in Fig.~\ref{bw11} corresponds to $\dot{D}$. We shall
return to these loops in Sec.~\ref{massless}.

\section{Symmetries derivable from the Poincar\'e Sphere}\label{poincs}

The Poincar\'e sphere serves as the basic language for polarization physics.
Its underlying language is the two-by-two coherency matrix.  This
coherency matrix contains the symmetry of $SL(2,c)$ isomorphic to the
the Lorentz group applicable to three space-like and one time-like
dimensions~\cite{hkn97,bk06jpa,kns13}.

\par

For polarized light propagating along the $z$ direction, the amplitude
ratio and phase difference of electric field $x$ and $y$
components traditionally determine the state of polarization. Hence,
the polarization can be changed by adjusting the amplitude
ratio or the  phase difference or both. Usually, the optical device which changes
amplitude is called an ``attenuator'' (or ``amplifier'') and the device which changes the relative
phase a ``phase shifter.''
\par
Let us start with the Jones vector:
\begin{equation}\label{jvec01}
\pmatrix{\psi_1(z,t) \cr \psi_2(z,t)} =
\pmatrix{a \exp{[i(kz - \omega t)]}  \cr a \exp{[i(kz - \omega t)] }} .
\end{equation}
To this matrix, we can apply the phase shift matrix of Eq.(\ref{rotz22})
which brings the Jones vector to
\begin{equation}\label{jvec03a}
\pmatrix{\psi_1(z,t) \cr \psi_2(z,t)} =
\pmatrix{a \exp{[i(kz - \omega t - i\phi/2)]}  \cr
a \exp{[i(kz - \omega t + i\phi/2)] }} .
\end{equation}
The generator of this phase-shifter is $J_3$ given Table~\ref{tab16}.
\par
The optical beam can be attenuated  differently in the two directions.
The resulting matrix is
\begin{equation} \label{jvec05a}
e^{-\mu}\pmatrix{e^{\eta/2} & 0 \cr 0 & e^{-\eta/2}}
\end{equation}
with the attenuation factor of $\exp{\left(-\mu_0 + \eta/2\right)}$
and $\exp{\left(-\mu - \eta/2\right)}$ for the $x$ and $y$ directions
respectively.  We are interested only the relative attenuation given
in Eq.(\ref{boostz22})
which leads to different amplitudes for the $x$ and $y$ component,
and the Jones vector becomes
\begin{equation}\label{jvec05c}
\pmatrix{\psi_1(z,t) \cr \psi_2(z,t)} =
\pmatrix{a e^{\mu/2} \exp{[i(kz - \omega t - i\phi/2)]}  \cr
a e^{-\mu/2}\exp{[i(kz - \omega t + i\phi/2)] }} .
\end{equation}
The squeeze matrix of Eq.(\ref{boostz22}) is generated by  $K_3$
given in Table~\ref{tab12}.

\par
The polarization is not always along the $x$ and $y$ axes, but
can be rotated around the $z$ axis using Eq.(\ref{rotz22})
generated by $J_2$ given in Table~\ref{tab12}.

\par
Among the rotation angles, the angle of $45^o$ plays an important role in
polarization optics.  Indeed, if we rotate the squeeze matrix of
Eq.(\ref{boostz22})
by $45^o$, we end up with the squeeze matrix of Eq.(\ref{boostx22})
generated by $K_1$ given also in Table~\ref{tab12}.
\par
Each of these four matrices plays an important role in special
relativity, as we discussed in Secs.~\ref{bwdecom} and \ref{isomor}.
Their respective roles
in optics and particle physics are given in Table~\ref{tab33}.
\par
The most general form for the two-by-two matrix applicable to the Jones
vector is the $G$ matrix of Eq.(\ref{alphabeta}).  This matrix is of course
a representation of the $SL(2,c)$ group.  It brings the simplest Jones
vector of Eq.(\ref{jvec01}) to its most general form.

%-------------------------------------------------------------------------------
\begin{table}%[ht]
\caption{Polarization optics and special relativity share the same mathematics.
Each matrix has its clear role in both optics and relativity.  The determinant
of the Stokes or the four-momentum matrix remains invariant under Lorentz
transformations.  It is interesting to note that the decoherence parameter
(least fundamental) in optics corresponds to the $(mass)^2$ (most fundamental) in
particle physics.}\label{tab33}
\vspace{2mm}
\begin{center}
\begin{tabular}{llcll}
\hline
\hline \\[0.5ex]
 Polarization Optics &\hspace{10mm}& Transformation Matrix  &\hspace{10mm} &
 Particle Symmetry \\[1.0ex]
\hline \\
Phase shift by $\phi$  &{}&
$\pmatrix{e^{-i\phi/2} & 0\cr 0 & e^{i\phi/2}}$
&{}&  Rotation around $z$.
\\[4ex]
Rotation around $z$  &{}&
$\pmatrix{\cos(\theta/2) & -\sin(\theta/2)\cr \sin(\theta/2) & \cos(\theta/2)}$
&{}&  Rotation around  $y$.
\\[4ex]
Squeeze along $x$ and $y$  &{}&
$\pmatrix{e^{\eta/2} & 0\cr 0 & e^{-\eta/2}}$
&{}&  Boost along $z$.
\\[4ex]
Squeeze along $45^o$  &{}&
$\pmatrix{\cosh(\lambda/2) & \sinh(\lambda/2)\cr \sinh(\lambda/2)
                & \cosh(\lambda/2)} $
&{}&   Boost along $x$.
\\[4ex]
a$^4$ $(\sin\xi)^2$ &{}& Determinant &{}&  (mass)$^2$
\\[4ex]
\hline
\hline\\[-0.8ex]
\end{tabular}
\end{center}
\end{table}
%-------------------------------------------------------------------------------------------

\subsection{Coherency Matrix}

However, the Jones vector alone cannot tell us whether the two components are
coherent with each other.  In order to address this important degree of
freedom, we use the coherency matrix defined as~\cite{born80,saleh07}
\begin{equation}\label{cocy11}
C = \pmatrix{S_{11} & S_{12} \cr S_{21} & S_{22}},
\end{equation}
where
\begin{equation}\label{cocy12}
<\psi_{i}^* \psi_{j}> = \frac{1}{T} \int_{0}^{T}\psi_{i}^* (t + \tau) \psi_{j}(t) dt,
\end{equation}
where $T$ is a sufficiently long time interval.
Then, those four elements become~\cite{hkn97}
\begin{eqnarray} \label{cocy15}
&{}& S_{11} = <\psi_{1}^{*}\psi_{1}> =a^2  , \qquad
S_{12} = <\psi_{1}^{*}\psi_{2}> = a^2(\cos\xi)e^{-i\phi} , \nonumber \\[1ex]
&{}& S_{21} = <\psi_{2}^{*}\psi_{1}> = a^2(\cos\xi)e^{+i\phi} ,  \qquad
S_{22} = <\psi_{2}^{*}\psi_{2}>  = a^2 .
\end{eqnarray}
The diagonal elements are the absolute values of $\psi_1$ and $\psi_2$
respectively.  The angle $\phi$ could be different from the value of
the phase-shift angle given in Eq.(\ref{rotz22}), but this difference
does not play any role in the reasoning.  The off-diagonal elements could be smaller than the product
of $\psi_1$ and $\psi_2$, if the two polarizations are not completely
coherent.
\par
The angle $\xi$ specifies the degree of coherency.  If it is zero, the
system is fully coherent, while the system is totally incoherent if $\xi$ is
$90^o$.   This can therefore be called the ``decoherence angle.''

\par
While the most general form of the transformation applicable to the
Jones vector is $G$ of Eq.(\ref{alphabeta}), the transformation
applicable to the coherency matrix is
\begin{equation}\label{cocy17}
           C' = G~C~G^{\dagger} .
\end{equation}
The determinant of the coherency matrix is invariant
under this transformation, and it is
\begin{equation}
\det(C) = a^4 (\sin\xi)^2 .
\end{equation}
Thus, angle $\xi$  remains invariant.  In the language
of the Lorentz transformation applicable to the four-vector, the
determinant is equivalent to the $(mass)^2$ and is therefore a Lorentz-invariant quantity.

\subsection{Two Radii of the Poincar\'e Sphere}\label{poincrad}

Let us write explicitly the transformation of Eq.(\ref{cocy17}) as
\begin{equation}\label{cocy21}
\pmatrix{S'_{11} & S'_{12} \cr S'_{21} & S'_{22}}
= \pmatrix{\alpha & \beta \cr \gamma & \delta}
\pmatrix{S_{11} & S_{12} \cr S_{21} & S_{22}}
\pmatrix{\alpha^{*} & \gamma^{*} \cr \beta^{*} & \delta^{*}} .
\end{equation}
It is then possible to construct the following quantities,
\begin{eqnarray}\label{stokes11}
&{}& S_{0} = \frac{S_{11} + S_{22}}{2},  \qquad
    S_{3} = \frac{S_{11} - S_{22}}{2}, \nonumber\\[2ex]
&{}& S_{1} = \frac{S_{12} + S_{21}}{2}, \qquad
 S_{2} = \frac{S_{12} - S_{21}}{2i}.
\end{eqnarray}
These are known as the Stokes parameters, and constitute a four-vector
$\left(S_0, S_3, S_1, S_2\right)$ under the Lorentz transformation.
\par
Returning to Eq.(\ref{rotz22}), the amplitudes of the two
orthogonal component are equal,  thus, the two diagonal elements of
the coherency matrix are equal.  This leads to $S_3 = 0$, and the
problem is reduced from the sphere to a circle.
\par

In this two-dimensional subspace, we can introduce the polar
coordinate system with
\begin{eqnarray}
&{}&  R = \sqrt{S_1^2 + S_2^2} \nonumber\\[1ex]
&{}&  S_1 = R \cos\phi, \nonumber\\[1ex]
&{}&  S_2 = R \sin\phi.
\end{eqnarray}
\par

%----------------------------------------------------------------------
\begin{figure}%[thb]
\centerline{\includegraphics[scale=0.3]{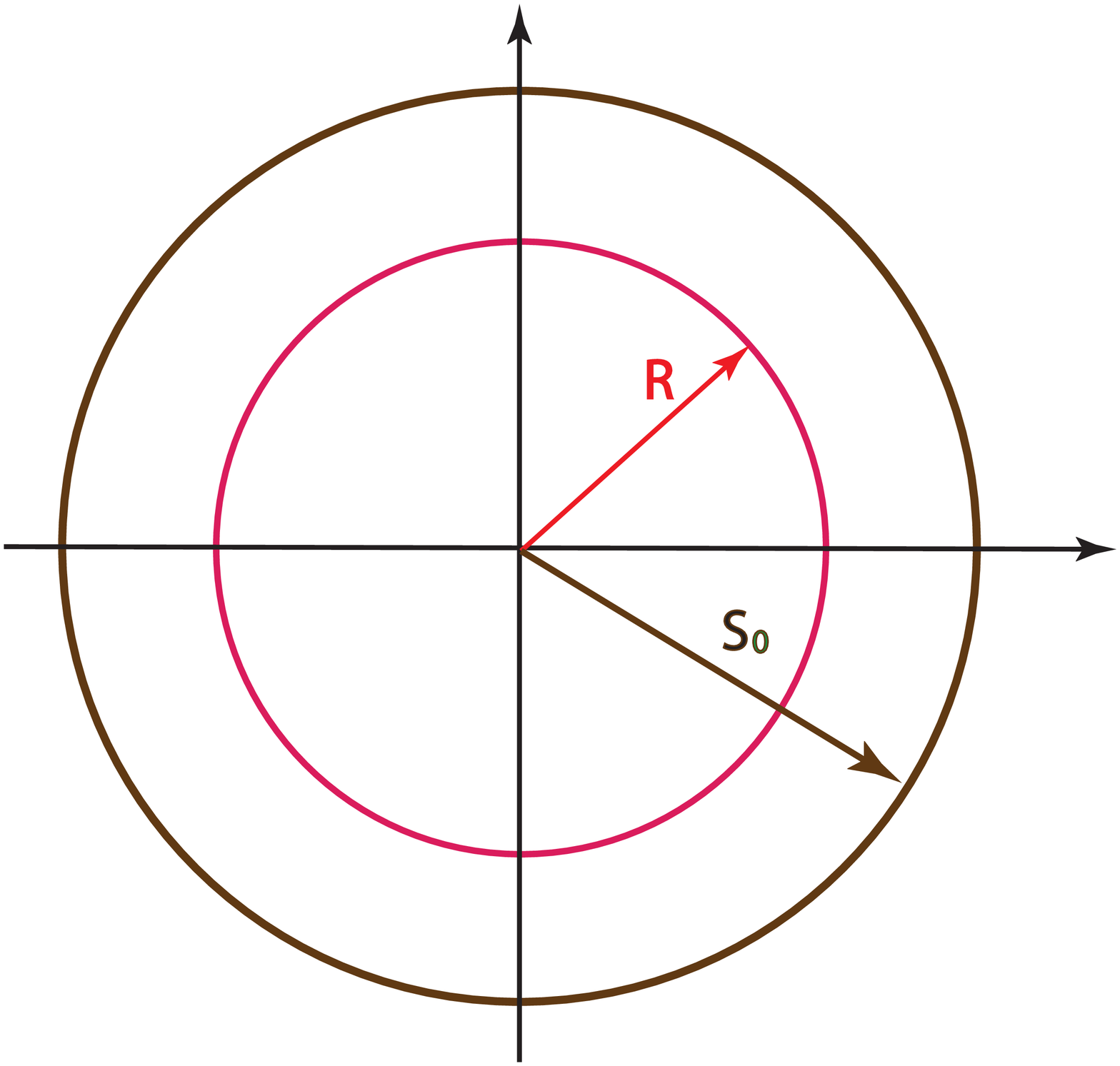}}
\caption{Radius of  the Poincar\'e sphere.  The radius $R$ takes its
maximum value $S_0$ when the decoherence angle $\xi$ is zero.  It becomes
smaller as $\xi$ increases.  It becomes zero when the angle reaches
$90^o$.}\label{poincs55}
\end{figure}
%----------------------------------------------------------------------

The radius $R$ is the radius of this circle, and is
\begin{equation}\label{radius11}
 R = a^2\cos\xi .
\end{equation}
\par
The radius $R$ takes its maximum value $S_0$ when $\xi = 0^o$.  It
decreases as $\xi$ increases and vanishes when $\xi = 90^o$.
This aspect of the radius R is illustrated in Fig.~\ref{poincs55}.
\par

%------------------------------------------------------------------------

In order to see its implications in special relativity,
let us go back to the four-momentum matrix of $m(1, 0, 0, 0)$.  Its
determinant is $m^2$ and remains invariant.   Likewise, the determinant
of the coherency matrix of Eq.(\ref{cocy11}) should also remain invariant.
The determinant in this case is
\begin{equation}
 S_{0}^2 - R^2 = a^4 \sin^2\xi .
\end{equation}
This quantity remains invariant under the Hermitian transformation of
Eq.(\ref{cocy21}), which is a Lorentz transformation as discussed
in Secs.~\ref{bwdecom} and \ref{isomor}.   This aspect is shown on the
last row of Table~\ref{tab33}.
\par

The coherency matrix then becomes
\begin{equation}\label{cocy55}
C = a^2 \pmatrix{1 & (\cos\xi)e^{-i\phi} \cr
(\cos\xi)e^{i\phi} & 1}.
\end{equation}
Since the angle $\phi$ does not play any essential role, we can let
$\phi = 0$, and write the coherency matrix as
\begin{equation}\label{cocy56}
C = a^2 \pmatrix{1 & \cos\xi \cr \cos\xi & 1}.
\end{equation}
The determinant of the above two-by-two matrix is
\begin{equation}
     a^4 \left(1 - \cos^2\xi\right) = a^4\sin^2\xi .
\end{equation}
\par
Since the Lorentz transformation leaves the determinant invariant, the
change in this $\xi$ variable is not a Lorentz transformation.  It is of
course possible to construct a larger group in which this variable plays
a role in a group transformation~\cite{bk06jpa}, but here we are
more interested in its role in a particle gaining a mass from zero or
the mass becoming zero.

%----------------------------------------------------------------------
\begin{figure}%[thb]
\centerline{\includegraphics[scale=0.6]{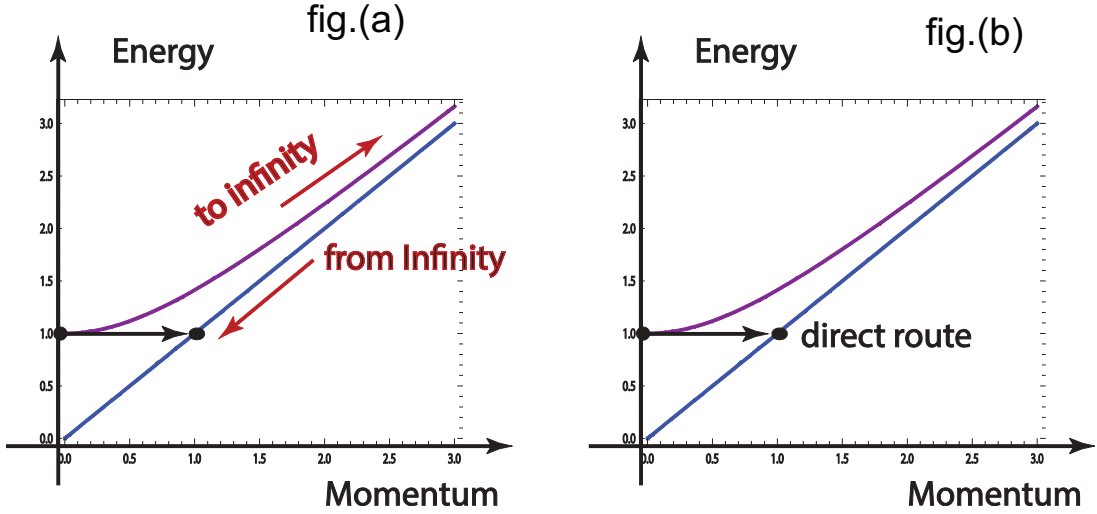}}
\caption{Transition from the massive to massless case.  Within the
framework of the Lorentz group, it is not possible to go from the massive
to massless case directly, because it requires the change in the mass
which is a Lorentz-invariant quantity.  The only way is to move to infinite
momentum and jump from the hyperbola to the light cone, and come back, as
illustrated in fig.(a).  The extra symmetry of the Poincar\'e sphere
allows a direct transition as shown in fig. (b).}\label{hypers}
\end{figure}
%----------------------------------------------------------------------

\subsection{Extra-Lorentzian Symmetry}\label{extralo}

The coherency matrix of Eq.(\ref{cocy55}) can be diagonalized to
\begin{equation}\label{cocy66}
a^2 \pmatrix{1 + \cos\xi & 0 \cr 0 & 1 - \cos\xi}
\end{equation}
by a rotation.
Let us then go back to the four-momentum matrix of Eq.(\ref{mom00}).  If
$p_x = p_y = 0$, and $p_z = p_{0} \cos\xi$ ,  we can write this matrix as
\begin{equation}\label{cocy68}
p_0 \pmatrix{1 + \cos\xi & 0 \cr 0 & 1 - \cos\xi} .
\end{equation}
\par
Thus, with this extra variable, it is possible to study the little groups
for variable masses, including the small-mass limit and the zero-mass case.
\par
For a fixed value of $p_0$, the $(mass)^2$ becomes
\begin{equation}\label{cocy70}
  (mass)^2 = \left(p_{0}\sin\xi\right)^2 , \quad\mbox{and}\quad
  (momentum)^2 = \left(p_{0}\cos\xi\right)^2 ,
\end{equation}
resulting in
\begin{equation}\label{cocy75}
(energy)^2 = (mass)^2 + (momentum)^2.
\end{equation}
\par
\begin{figure}%[thb]
\centerline{\includegraphics[scale=0.3]{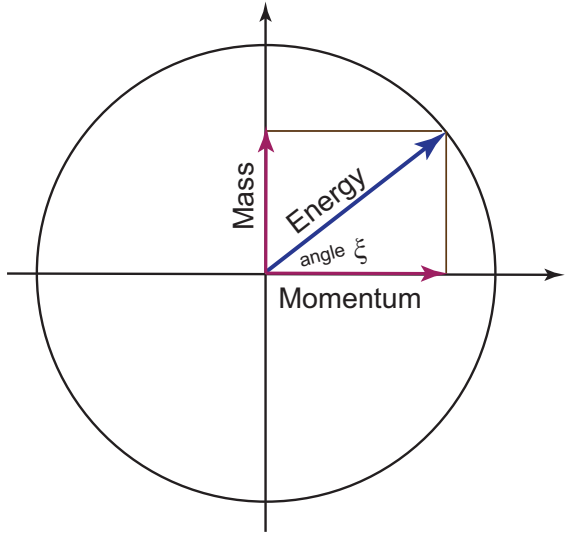}}
\caption{Energy-momentum-mass relation.  This circle illustrates the case
where the energy is fixed, while the mass and momentum are related according
to the triangular rule.  The value of the angle $\xi$ changes from zero to
$180^{o}$.  The particle mass is negative for negative values of this angle.
However, in the Lorentz group, only $(mass)^2$ is a relevant variable, and
negative masses might play a role for theoretical purposes. }\label{extrav}
\end{figure}

This transition is illustrated in Fig.~\ref{hypers}.
We are interested in reaching a point on the light cone from mass hyperbola
while keeping the energy fixed.   According to this figure, we do not
have to make an excursion to infinite-momentum limit.  If the energy is
fixed during this process, Eq.(\ref{cocy75}) tells the mass and momentum
relation, and Figure~\ref{extrav} illustrates this relation.

\par
Within the framework of the Lorentz group, it is possible, by making
an excursion to  infinite momentum where the mass hyperbola coincides
with the light cone, to then come back to the desired point.  On the
other hand, the mass formula of Eq.(\ref{cocy70}) allows us to go there
directly.  The decoherence mechanism of the coherency matrix makes this
possible.

%----------------------------------------------------------------------
%----------------------------------------------------------------------

\section{Small-mass and Massless Particles}\label{massless}

We now have a mathematical tool to reduce the mass of a massive
particle from its positive value to zero.  During this process,
the Lorentz-boosted rotation matrix becomes a gauge transformation
for the spin-1 particle, as discussed Subsec.~\ref{e2symm}.
For spin-1/2 particles, there are two issues.

\begin{itemize}

\item[1.]It was seen in Subsec.~\ref{e2symm} that the requirement of
gauge invariance lead to a polarization of massless spin-1/2 particle,
such as neutrinos.   What happens to anti-neutrinos?

\item[2.]There are strong experimental indications that neutrinos
have a small mass.  What happens to the $E(2)$ symmetry?

\end{itemize}

\subsection{Spin-1/2 Particles}\label{spinhalf}

Let us go back to the two-by-two matrices of Subsec.~\ref{conju},
and the two-by-two $D$ matrix.  For a massive particle, its
Wigner decomposition leads to
\begin{equation}\label{sm01}
D = \pmatrix{\cos(\theta/2) & -e^{-\eta}\sin(\theta/2) \cr
              e^{\eta}\sin(\theta/2) & \cos(\theta/2)}
\end{equation}
This matrix is applicable to the spinors $u$ and $v$ defined in Eq.(\ref{e242})
respectively for the spin-up and spin-down states along the
$z$ direction.

Since the Lie algebra of $SL(2,c)$ is invariant under the
sign change of the $K_i$ matrices, we can consider
the ``dotted'' representation, where the system is boosted
in the opposite direction, while the direction of rotations
remain the same.  Thus, the Wigner decomposition leads to
\begin{equation}\label{sm05}
\dot{D} = \pmatrix{\cos(\theta/2) & -e^{\eta}\sin(\theta/2) \cr
              e^{-\eta}\sin(\theta/2) & \cos(\theta/2)}
\end{equation}
with its spinors
\begin{equation}\label{sm07}
\dot{u} = \pmatrix{1 \cr 0}, \quad \mbox{and}\quad
  \dot{v} = \pmatrix{0 \cr 1}.
\end{equation}
\par

For anti-neutrinos, the helicity is reversed but the momentum is
unchanged.  Thus, $D^{\dagger}$ is the appropriate matrix.  However,
$ D^{\dagger} = \dot{D}^{-1} $ as was noted in Subsec~\ref{conju}.  Thus, we shall use $\dot{D}$
for anti-neutrinos.

\par

When the particle mass becomes very small,
\begin{equation}
e^{-\eta} = \frac{m}{2p} ,
\end{equation}
becomes small.  Thus, if we let
\begin{equation}
e^{\eta} \sin(\theta/2) = \gamma, \quad\mbox{and}\quad
            e^{-\eta}\sin(\theta/2) = \epsilon^2 ,
\end{equation}
then the $D$  matrix of Eq.(\ref{sm01}) and the $\dot{D}$ of Eq.(\ref{sm05}) become
\begin{equation}
\pmatrix{1 - \gamma \epsilon^2/2  &  -\epsilon^2 \cr
     \gamma & 1 - \gamma\epsilon^2},  \quad\mbox{and}\quad
\pmatrix{1 - \gamma \epsilon^2/2  &  -\gamma \cr
  \epsilon^2 & 1 - \gamma\epsilon^2},
\end{equation}
respectively where $\gamma$ is an independent parameter and
\begin{equation}
\epsilon^2 = \gamma \left(\frac{m}{2p}\right)^2 .
\end{equation}
\par
When the particle mass becomes zero, they become
\begin{equation}\label{sm16}
\pmatrix{1  & 0 \cr \gamma & 1},
\quad\mbox{and}\quad \pmatrix{1  &  -\gamma \cr 0 & 1},
\end{equation}
respectively, applicable to the spinors $(u, v)$ and $(\dot{u},\dot{v})$
respectively.
\par
For neutrinos,
\begin{equation}\label{sm18}
\pmatrix{ 1 & 0 \cr \gamma & 1}\pmatrix{1 \cr 0} = \pmatrix{1 \cr \gamma},
\quad\mbox{and}\quad
\pmatrix{ 1 & 0 \cr \gamma & 1}\pmatrix{0 \cr 1} = \pmatrix{0 \cr 1} .
\end{equation}
For anti-neutrinos,
\begin{equation}\label{sm20}
\pmatrix{ 1 & -\gamma \cr 0 & 1}\pmatrix{1 \cr 0} = \pmatrix{1 \cr 0},
\quad\mbox{and}\quad
\pmatrix{ 1 & -\gamma \cr 0 & 1}\pmatrix{0 \cr 1} = \pmatrix{-\gamma \cr 1}.
\end{equation}
\par
It was noted in Subsec.~\ref{e2symm} that the triangular matrices of
Eq.(\ref{sm16}) perform gauge transformations. Thus, for Eq.(\ref{sm18}) and
Eq.(\ref{sm20}) the requirement of gauge invariance leads to the polarization
of neutrinos.  The neutrinos are left-handed while the anti-neutrinos are
right-handed.  Since, however, nature cannot tell the difference
between the dotted and undotted representations, the Lorentz group cannot
tell which neutrino is right handed. It can say only that the neutrinos
and anti-neutrinos are oppositely polarized.
\par
If the neutrino has a small mass, the gauge invariance is
modified to

\begin{equation}
\pmatrix{1 - \gamma \epsilon^2/2  &  -\epsilon^2 \cr
     \gamma & 1 - \gamma\epsilon^2/2}
    \pmatrix{0 \cr 1} = \pmatrix{0 \cr 1} -
      \epsilon^2 \pmatrix{1 \cr  \gamma/2} ,
\end{equation}
and
\begin{equation}
\pmatrix{1 - \gamma \epsilon^2/2  &  -\gamma \cr
  \epsilon^2 & 1 - \gamma\epsilon^2} \pmatrix{1 \cr 0}
  = \pmatrix{1 \cr 0} + \epsilon^2 \pmatrix{-\gamma/2 \cr  1},
\end{equation}
respectively for neutrinos and anti-neutrinos. Thus the violation of the
gauge invariance in both cases is proportional to $\epsilon^2$ which is
$m^2/4p^2$.

\subsection{Small-mass neutrinos in the Real World}

Whether neutrinos have mass or not and the consequences of this relative
to the Standard Model and lepton number is the subject of much theoretical
speculation~\cite{papoulias_2013,dinh_2013}, and of cosmology~\cite{miramonti_2013},
nuclear reactors~\cite{sinev_2013, li_2013}, and high energy
experimentations~\cite{bergstrom_2013,han_2013}.
Neutrinos are fast becoming an important component of the search for
dark matter and dark radiation~\cite{drewes_2013}.  Their importance
within the Standard Model is reflected by the fact that they are the
only particles which seem to exist with only one direction of chirality,
i.e. only left-handed neutrinos have been confirmed to exist so far.

\par
It was speculated some time ago that neutrinos in constant electric and
magnetic fields would acquire a small mass, and that right-handed neutrinos
would be trapped within the interaction field~\cite{barut_1986}. Solving
generalized electroweak models using left- and right-handed neutrinos
has been discussed recently~\cite{palcu_2006}.   Today these
right-handed neutrinos which do not participate in weak interactions are
called ``sterile'' neutrinos~\cite{bilenky_2013}.  A comprehensive
discussion of the place of neutrinos in the scheme of physics has been
given by Drewes~\cite{drewes_2013}.

\section{Scalars, Four-vectors, and Four-Tensors}\label{tensor}

In Secs.~\ref{complete}  and \ref{massless}, our primary interest has been
the two-by-two matrices applicable to spinors for spin-1/2 particles.
Since we also used four-by-four matrices, we indirectly studied the
four-component particle consisting of spin-1 and spin-zero components.
\par
If there are two spin 1/2 states, we are accustomed to construct one
spin-zero state, and one spin-one state with three degeneracies.
\par
In this paper, we are confronted with two spinors, but each spinor can also
be dotted.  For this reason, there are sixteen orthogonal states
consisting of spin-one and spin-zero states.  How many spin-zero states?
How many spin-one states?

 \par

For particles at rest, it is known that the addition of two one-half spins
result in spin-zero and spin-one states.  In the this paper, we have two
different spinors behaving differently under the Lorentz boost.
Around the $z$ direction, both spinors are transformed by
\begin{equation} \label{max01}
Z(\phi) = \exp{\left(-i\phi J_3\right)}
= \pmatrix{e^{-i\phi/2} & 0 \cr 0 & e^{i\phi/2}}.
\end{equation}
However, they are boosted by
\begin{eqnarray} \label{max03}
&{}& B(\eta) = \exp{\left(-i\eta K_3\right)}
= \pmatrix{e^{\eta/2} & 0 \cr  0 & e^{-\eta/2}} , \nonumber\\[2ex]
&{}& \dot{B}(\eta) = \exp{\left(i\eta K_3\right)} ,
= \pmatrix{e^{-\eta/2} & 0 \cr  0 & e^{\eta/2}}
\end{eqnarray}
applicable to the undotted and dotted spinors respectively.  These two
matrices commute with each other, and also with the rotation matrix
$Z(\phi)$ of Eq.(\ref{max01}).  Since $K_3$ and $J_3$ commute with each
other, we can work with the matrix $Q(\eta,\phi)$ defined as
\begin{eqnarray}\label{max05}
&{}& Q(\eta,\phi) = B(\eta)Z(\phi) = \pmatrix{e^{(\eta - i\phi)/2} &  0 \cr
  0 & e^{-(\eta - i\phi)/2}} ,  \nonumber \\[2ex]
&{}& \dot{Q}(\eta, \phi) = \dot{B}(\eta) \dot{Z}(\phi)
= \pmatrix{e^{-(\eta + i\phi)/2} &  0 \cr
  0 & e^{(\eta + i\phi)/2}} .
\end{eqnarray}

\par
When this combined matrix is applied to the spinors,
\begin{eqnarray}\label{max07}
&{}& Q(\eta,\phi) u = e^{(\eta -i\phi)/2} u , \qquad
Q(\eta,\phi) v = e^{-(\eta -i\phi)/2} v,
\nonumber\\[2ex]
&{}& \dot{Q}(\eta, \phi) \dot{u} = e^{-(\eta + i\phi)/2} \dot u , \qquad
  \dot{Q}(\eta,\phi)\dot{v} = e^{(\eta + i\phi)/2} \dot{v}.
\end{eqnarray}

\par

If the particle is at rest, we can construct the combinations
\begin{equation}\label{max09}
 uu, \qquad \frac{1}{\sqrt{2}}(uv + vu),
\qquad vv,
\end{equation}
to construct the spin-1 state, and
\begin{equation} \label{max10}
\frac{1}{\sqrt{2}}(uv - vu) ,
\end{equation}
for the spin-zero state.  There are four bilinear states.
In the $SL(2,c)$ regime, there are two dotted spinors.   If we
include both dotted and undotted spinors, there are sixteen independent
bilinear combinations.  They are given in Table~\ref{tab51}.  This
table also gives the effect of the operation of $Q(\eta, \phi)$.

%-----------------------------------------------------------------------
\begin{table}
\caption{Sixteen combinations of the $SL(2,c)$ spinors.  In the $SU(2)$
regime, there are two spinors leading to four bilinear forms.  In
the $SL(2,c)$ world, there are two undotted and two dotted spinors.
These four spinors lead to sixteen independent bilinear combinations.}\label{tab51}
\begin{center}
\begin{tabular}{ccccc}
\hline \\[-2.4ex]
\hline \\
{}& Spin 1 & \hspace{5mm} & Spin 0 & {} \\[2ex]
\hline  \\
{}& $ uu,\quad \frac{1}{\sqrt{2}}(uv + vu), \quad  vv, $   & {} &
$\frac{1}{\sqrt{2}}(uv - vu) $   &  {} \\[2ex]
\hline  \\
{}& $ \dot{u}\dot{u},\quad \frac{1}{\sqrt{2}}(\dot{u}\dot{v} + \dot{v}\dot{u}),
  \quad  \dot{v}\dot{v}, $   & {} &
$\frac{1}{\sqrt{2}}(\dot{u}\dot{v} - \dot{v}\dot{u}) $   &  {} \\[2ex]
\hline  \\
{}& $ u\dot{u},\quad \frac{1}{\sqrt{2}}(u\dot{v} + v\dot{u}),
\quad  v\dot{v}, $   & {} &
$\frac{1}{\sqrt{2}}(u\dot{v} - v\dot{u}) $   &  {} \\[2ex]
\hline  \\
{}& $\dot{u}u, \quad \frac{1}{\sqrt{2}}(\dot{u}v + \dot{v}u),
\quad  \dot{v}v, $   & {} &
$\frac{1}{\sqrt{2}}( \dot{u}v -  \dot{v}u) $   &  {} \\[2ex]
\hline\\[-2.4ex]
%-----------------------------------------------------------------
%-----------------------------------------------------------------
%-----------------------------------------------------------------
\hline \\
{}& After the operation of $Q(\eta,\phi)$ and $\dot{Q}(\eta,\phi)$
  & \hspace{10mm} &  & {} \\[2ex]
\hline  \\
{}& $ e^{-i\phi} e^{\eta} uu,\quad \frac{1}{\sqrt{2}}(uv + vu),
\quad  e^{i\phi}e^{-\eta}  vv,
$ & {} &
$\frac{1}{\sqrt{2}}(uv - vu) $   &  {} \\[2ex]
\hline  \\
{}& $ e^{-i\phi}e^{-\eta}  \dot{u}\dot{u},\quad
\frac{1}{\sqrt{2}}(\dot{u}\dot{v} +\dot{v}\dot{u}),
  \quad e^{i\phi}e^{\eta}   \dot{v}\dot{v}, $   & {} &
$\frac{1}{\sqrt{2}}(\dot{u}\dot{v} - \dot{v}\dot{u}) $   &  {} \\[2ex]
\hline  \\
{}& $ e^{-i\phi} u\dot{u},\quad
\frac{1}{\sqrt{2}}(e^{\eta} u\dot{v} + e^{-\eta} v\dot{u}),
\quad  e^{i\phi} v\dot{v}, $   & {} &
$\frac{1}{\sqrt{2}}(e^{\eta} u\dot{v} - e^{-\eta} v\dot{u}) $   &  {} \\[2ex]
\hline  \\
{}& $e^{-i\phi} \dot{u}u, \quad \frac{1}{\sqrt{2}}(\dot{u}v + \dot{v}u),
\quad  e^{i\phi} \dot{v}v, $   & {} &
$\frac{1}{\sqrt{2}}(e^{-\eta} \dot{u}v - e^{\eta} \dot{v}u) $   &  {} \\[2ex]
\hline\\[-2.4ex]
\hline
\end{tabular}
\end{center}
\end{table}
%-----------------------------------------------------------------------

\par
Among the bilinear combinations given in Table~\ref{tab51}, the following two
are invariant under rotations and also under boosts.
\begin{equation}\label{max11}
S =   \frac{1}{\sqrt{2}}(uv - vu), \quad\mbox{and}\quad
\dot{S} = - \frac{1}{\sqrt{2}}(\dot{u}\dot{v} - \dot{v}\dot{u}) .
\end{equation}
They are thus scalars in the Lorentz-covariant world.  Are they the same
or different?  Let us consider the following combinations
\par
\begin{equation}\label{max12}
S_{+} = \frac{1}{\sqrt{2}}\left(S + \dot{S}\right), \quad\mbox{and}\quad
S_{-} = \frac{1}{\sqrt{2}}\left(S - \dot{S}\right) .
\end{equation}
Under the dot conjugation, $S_{+}$ remains invariant, but $S_{-}$ changes
its sign.
\par
Under the dot conjugation, the boost is performed in the opposite
direction.  Therefore it is the operation of space inversion, and
$S_{+}$ is a scalar while $S_{-}$ is called the pseudo-scalar.

\subsection{Four-vectors} \label{4vec}

Let us consider the bilinear products of one dotted and one undotted
spinor as~ $ u\dot{u},~ u\dot{v},~ \dot{u}v,
~ v\dot{v}$, and construct the matrix
\begin{equation}\label{max15}
U = \pmatrix{u \dot{v} & v \dot{v} \cr u \dot{u} & v\dot{u} }.
\end{equation}
Under the rotation $Z(\phi)$ and the boost $B(\eta)$ they become
\begin{equation}\label{max17}
 \pmatrix{e^{\eta} u \dot{v} &  e^{-i\phi} v \dot{v} \cr
 e^{i\phi} u \dot{u} & e^{-\eta} v \dot{u}}.
\end{equation}
Indeed, this matrix is consistent with the transformation properties
given in Table~\ref{tab51}, and
transforms like the four-vector
\begin{equation}\label{max19}
\pmatrix{ t + z & x - iy \cr x + iy & t - z }.
\end{equation}
This form was given in Eq.(\ref{2b2}),  and played the central role
throughout this paper.  Under the space inversion, this
matrix becomes
\begin{equation}\label{max20}
\pmatrix{ t - z &  -(x - iy) \cr -(x + iy) & t + z }.
\end{equation}
This space inversion is known as the parity operation.

\par
The form of Eq.(\ref{max15}) for a particle or field with
four-components, is given by
$(V_{0}, V_{z}, V_{x}, V_{y})$.  The two-by-two form of this four-vector is
\begin{equation}\label{max21}
U = \pmatrix{V_{0} + V_{z} & V_{x} - iV_{y} \cr
V_{x} + iV_{y} & V_{0} - V_{z}}.
\end{equation}
\par

If boosted along the $z$ direction, this matrix becomes
\begin{equation}\label{max22}
\pmatrix{e^{\eta}\left(V_{0} + V_{z}\right) & V_{x} - iV_{y} \cr
V_{x} + iV_{y} & e^{-\eta}\left(V_{0} - V_{z}\right)}.
\end{equation}
\par

In the mass-zero limit, the four-vector matrix of Eq.(\ref{max22})
becomes
\begin{equation} \label{max23}
\pmatrix{ 2A_{0} & A_{x} - iA_{y} \cr
A_{x} + iA_{y} &  0 },
\end{equation}
with the Lorentz condition $A_{0} = A_{z}$. The gauge
transformation applicable to the photon four-vector was
discussed in detail in Subsec.~\ref{e2symm}.
\par

Let us go back to the matrix of Eq.(\ref{max21}), we can construct
another matrix $\dot{U}$.  Since the dot conjugation leads to
the space inversion,
\begin{equation}\label{max25}
\dot{U} = \pmatrix{\dot{u}v & \dot{v} v \cr
              \dot{u} u & \dot{v}u }.
\end{equation}
Then
\begin{eqnarray}\label{max27}
&{}& \dot{u}v \simeq (t - z), \qquad \dot{v}u  \simeq  (t + z)
\nonumber\\[1ex]
&{}& \dot{v}v \simeq -(x - iy), \qquad \dot{u}u  \simeq -(x + iy) ,
\end{eqnarray}
where the symbol $\simeq$ means ``transforms like.''
\par
Thus, $U$ of Eq.(\ref{max15}) and $\dot{U}$ of Eq.(\ref{max25}) used up
eight of the sixteen bilinear forms.  Since there are two bilinear forms
in the scalar and pseudo-scalar as given in Eq.(\ref{max12}), we have
to give interpretations to the six remaining bilinear forms.

\subsection{Second-rank Tensor}

In this subsection, we are studying bilinear forms with both spinors dotted
and undotted. In Subsec.~\ref{4vec}, each bilinear spinor consisted of one
dotted and one undotted spinor.   There are also bilinear spinors
which are both
dotted  or both undotted.  We are interested in two sets of three quantities
satisfying the $O(3)$ symmetry. They should therefore transform like
\begin{equation}
(x + iy)/\sqrt{2}, \qquad (x - iy)/\sqrt{2}, \qquad  z ,
\end{equation}
which are like
\begin{equation}
  uu, \qquad vv,  \quad (uv + vu)/\sqrt{2} ,
\end{equation}
respectively in the $O(3)$ regime.  Since the dot conjugation is
the parity operation, they are like
\begin{equation}
-\dot{u}\dot{u}, \qquad -\dot{v}\dot{v}, \qquad
-(\dot{u}\dot{v} + \dot{v}\dot{u})/\sqrt{2} .
\end{equation}
In other words,
\begin{equation}
(uu{\dot )} = -\dot{u}\dot{u}, \quad\mbox{and}\quad
                    (vv{\dot)} = -\dot{v}\dot{v}.
 \end{equation}
We noticed a similar sign change in Eq.(\ref{max27}).

\par

In order to construct the $z$ component in this $O(3)$ space, let us first
consider
\begin{equation}\label{max51}
f_{z} = \frac{1}{2}\left[(uv + vu) -
   \left(\dot{u}\dot{v} + \dot{v}\dot{u}\right)\right] ,
\qquad
g_{z} = \frac{1}{2i}\left[(uv + vu) +
\left(\dot{u}\dot{v} + \dot{v}\dot{u}\right)\right] ,
\end{equation}
where $f_{z}$ and $g_{z}$ are respectively symmetric and anti-symmetric
under the dot conjugation or the parity operation.  These quantities
are invariant under the boost along the $z$ direction. They are also
invariant under rotations around this axis, but they are not invariant
under boost along or rotations around the $x$ or $y$ axis.  They are
different from the scalars given in Eq.(\ref{max11}).
\par
Next, in order to construct the $x$ and $y$ components, we start with
$g_{\pm}$ as
\begin{eqnarray}\label{max53}
&{}& f_{+} = \frac{1}{\sqrt{2}}\left(uu - \dot{u}\dot{u}\right)
   \qquad
 g_{+} = \frac{1}{\sqrt{2}i}\left(uu + \dot{u}\dot{u}\right)
 \nonumber\\[1ex]
&{}& f_{-} = \frac{1}{\sqrt{2}}\left(vv - \dot{v}\dot{v}\right)
   \qquad
 g_{-} = \frac{1}{\sqrt{2}i}\left(vv + \dot{v}\dot{v}\right).
\end{eqnarray}
Then
\begin{eqnarray}\label{max55}
&{}& f_x = \frac{1}{\sqrt{2}}\left(f_{+} + f_{-}\right)
= \frac{1}{2}\left[ \left(uu - \dot{u}\dot{u}\right)
 + \left(vv - \dot{v}\dot{v}\right)\right]  \nonumber\\[1ex]
&{}& f_y = \frac{1}{\sqrt{2}i}\left(f_{+} - f_{-}\right)
= \frac{1}{2i}\left[ \left(uu - \dot{u}\dot{u}\right)
- \left(vv -  \dot{v}\dot{v}\right)\right] .
\end{eqnarray}
and
\begin{eqnarray}\label{max57}
&{}& g_{x} = \frac{1}{\sqrt{2}}\left(g_{+} + g_{-}\right)
= \frac{1}{2i}\left[ \left(uu  + \dot{u}\dot{u}\right)
 + \left(vv + \dot{v}\dot{v}\right)\right]  \nonumber\\[1ex]
&{}& g_{y} = \frac{1}{\sqrt{2}i}\left(g_{+} - g_{-}\right)
= -\frac{1}{2}\left[ \left(uu + \dot{u}\dot{u}\right)
- \left(vv + \dot{v}\dot{v}\right)\right] .
\end{eqnarray}
Here  $f_{x}$ and  $f_{y}$ are symmetric under dot conjugation, while
$g_{x}$ and  $g_{y}$ are anti-symmetric.

\par
Furthermore, $f_{z}, f_{x},$ and  $f_{y}$  of Eqs.(\ref{max51}) and
(\ref{max55}) transform like a  three-dimensional vector. The same can be
said for $g_{i}$ of Eqs.(\ref{max51}) and (\ref{max57}).   Thus, they can
grouped into the second-rank tensor
\begin{equation}\label{max59}
T = \pmatrix{
    0 & -g_z & -g_x &  -g_y \cr g_z & 0 & -f_y &  f_x \cr
    g_x  & f_y & 0 &  -f_z \cr  g_y &  -f_x & f_z & 0
    } ,
\end{equation}
whose Lorentz-transformation properties are well known.  The
$g_{i}$ components change their signs under space inversion,
while the $f_{i}$ components remain invariant.  They are like
the electric and magnetic fields respectively.

\par

If the system is Lorentz-booted, $f_{i}$ and $g_{i}$ can be computed from
Table~\ref{tab51}.  We are now interested in the symmetry of photons by
taking the massless limit.  According to the procedure developed in
Sec.~\ref{poincs}, we can keep only the terms which become larger for
larger values of $\eta$. Thus,
\begin{eqnarray}\label{max58}
&{}&  f_{x} \rightarrow \frac{1}{2} \left(uu - \dot{v}\dot{v}\right), \qquad
  f_{y} \rightarrow \frac{1}{2i} \left(uu + \dot{v}\dot{v}\right),
  \nonumber\\[2ex]
&{}& g_{x} \rightarrow \frac{1}{2i} \left(uu + \dot{v}\dot{v}\right),
 \qquad
  g_{y} \rightarrow -\frac{1}{2} \left(uu  - \dot{v}\dot{v}\right) ,
\end{eqnarray}
in the massless limit.

\par

Then the tensor of Eq.(\ref{max59}) becomes
\begin{equation}\label{max52}
F = \pmatrix{ 0  & 0 & -E_{x}  & -E_{y} \cr
0  & 0 & -B_{y} &  B_{x} \cr
E_{x} & B_{y} & 0  &  0  \cr
E_{y} & -B_{x}  & 0 &   0 } ,
\end{equation}
with
\begin{eqnarray}\label{max60}
&{}&  B_{x} \simeq \frac{1}{2} \left(uu - \dot{v}\dot{v}\right),
  \qquad
  B_{y} \simeq \frac{1}{2i} \left(uu + \dot{v}\dot{v}\right),
  \nonumber\\[2ex]
&{}& E_{x} = \frac{1}{2i} \left(uu + \dot{v}\dot{v}\right),
 \qquad
  E_{y} = -\frac{1}{2} \left(uu - \dot{v}\dot{v}\right) .
\end{eqnarray}

\par

The electric and magnetic field components are perpendicular to
each other.  Furthermore,
\begin{equation}
E_{x} = B_{y}, \quad\mbox\quad E_{y} = -B_{x} .
\end{equation}

\par

In order to address this question, let us go back to Eq.(\ref{max53}).
In the massless limit,
\begin{equation}\label{max63}
B_{+} \simeq E_{+} \simeq uu, \qquad B_{-} \simeq E_{-}
    \simeq \dot{v}\dot{v}
\end{equation}
The gauge transformation applicable to $u$ and $\dot{v}$ are
the two-by-two matrices
\begin{equation}
\pmatrix{1 & -\gamma \cr 0 & 1}, \quad\mbox{and}\quad
\pmatrix{1 & 0 \cr -\gamma & 1} .
\end{equation}
respectively as noted in Subsecs.~\ref{e2symm} and \ref{spinhalf}.
Both $u$ and $\dot{v}$ are invariant under gauge transformations,
while $\dot{u}$ and $v$ do not.
\par
The $B_{+}$ and $E_{+}$ are for the photon spin along the $z$ direction,
while  $B_{-}$ and $E_{-}$ are for the opposite direction.
In 1964~\cite{wein64b}, Weinberg constructed gauge-invariant state vectors
for massless particles starting from Wigner's 1939 paper~\cite{wig39}.
The bilinear spinors $uu$ and  and $\dot{v}\dot{v}$ correspond to Weinberg's
state vectors.

%----------------------------------------------------------------------
\begin{figure}[thb]
\centerline{\includegraphics[scale=0.5]{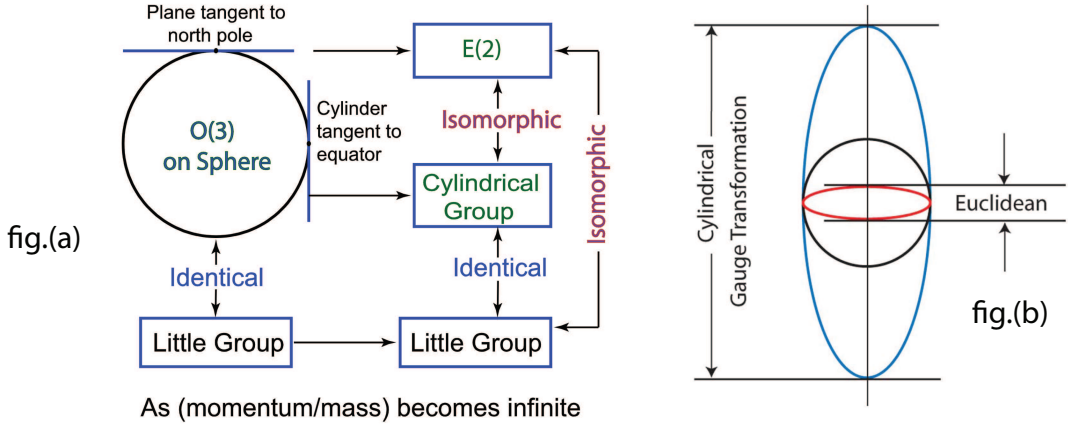}}
\caption{Contractions of the three-dimensional rotation group.  This
group can be illustrated by a sphere.  This group can become the
two-dimensional Euclidean group on a plane tangent at the north pole as
illustrated in fig.(a).  It was later noted that there is a cylinder
tangential to this sphere, and the up and down translations on this cylinder
correspond to the gauge transformation for photons~\cite{kiwi87jmp}.
As illustrated in fig.(b), the four-dimensional representation of
of the Lorentz group contains both elongation and contraction of
of the $z$ axis, as the system is boosted along this direction.
The elongation and the contraction become the cylindrical and
Euclidean groups, respectively~\cite{kiwi90jmp}.}\label{isomor}
\end{figure}
%----------------------------------------------------------------------

\subsection{Possible Symmetry of the Higgs Mechanism}

In this section, we discussed how the two-by-two formalism of the group
$SL(2,c)$ leads the scalar, four-vector, and tensor representations of the
Lorentz group.  We discussed in detail how the
four-vector for a massive particle can be decomposed into the symmetry of a
two-component massless particle and one gauge degree of freedom. This aspect
was studied in detail by Kim and Wigner~\cite{kiwi87jmp,kiwi90jmp}, and
their results are illustrated in Fig.~\ref{isomor}.

\par
This subject was initiated by In{\"o}n{\"u} and Wigner in 1953 as the group
contraction~\cite{inonu53}. In their paper, they discussed the contraction
of the three-dimensional rotation group becoming contracted to the
two-dimensional Euclidean group with one rotational and two translational
degrees of freedom.  While the $O(3)$ rotation group can be illustrated by
a three-dimensional sphere, the plane tangential at the north pole is for
the $E(2)$ Euclidean group. However, we can also consider a cylinder
tangential at the equatorial belt.  The resulting cylindrical group is
isomorphic to the Euclidean group~\cite{kiwi87jmp}.  While the rotational
degree of freedom of this cylinder is for the photon spin, the up and down translations
on the surface of the cylinder correspond to the gauge degree of freedom
of the photon, as illustrated in Fig.~\ref{isomor}.

\par
The four-dimensional Lorentz group contains both the Euclidean and
cylindrical contractions. These contraction processes transform a
four-component massive vector meson into a massless spin-one particle
with two spin one-half components, and one gauge degree of freedom.

\par
Since this contraction procedure is spelled out detail in
Ref.~\cite{kiwi90jmp}, as well as in the present paper, its reverse process
is also well understood.  We start with one two-component massless
particle with one gauge degree of freedom, and end up with a massive
vector meson with its four components.

\par
The mathematics of this process is not unlike the Higgs
mechanism~\cite{higgs64,kibble64}, where one massless field with two
degrees of freedom absorbs one gauge degree freedom to become a
quartet of bosons, namely that of $W, Z^{\pm}$ plus the Higgs boson.
As is well known, this mechanism is the basis for the theory of
electro-weak interaction formulated by Weinberg and
Salam~\cite{wein67,wein96}.

\par
The word "spontaneous symmetry breaking" is used for the Higgs mechanism.
It could be an interesting problem to see that this symmetry breaking for
the two Higgs doublet model can be formulated in terms of the Lorentz group and its contractions.
In this connection, we note an interesting recent paper by D{\'e}e and
Ivanov~\cite{degee10}.

\par

\section*{Conclusions}

It was noted in this paper that the second-order differential equation for
damped harmonic oscillators can be formulated in terms of two-by-two
matrices.  These matrices produce the algebra of the group $Sp(2)$.
While there are three trace classes of the two-by-two matrices of this
group, the damped oscillator tells us how to make transitions from one
class to another.
\par
It is shown that Wigner's three little groups can be defined in terms of
the trace classes of the $Sp(2)$ group.  If the trace is smaller than two,
the little group is for massive particles.  If greater than two, the
little group is for imaginary-mass particles.  If the trace is equal to
two, the little group is for massless particles.  Thus, the damped harmonic
oscillator provides a procedure for transition from one little group to
another.

\par
The Poincar\'e sphere contains the symmetry of the six-parameter
$SL(2,c)$group.  Thus, the sphere provides the procedure for extending
the symmetry of the little group defined within the space of three-dimensional
Lorentz group to the full four-dimensional Minkowski space. In addition,
the Poincar\'e sphere offers the variable which allows us to change the
symmetry of massive particle to that of massless particle by continuously
changing the mass.
\par
In this paper, we extracted the mathematical properties of the Lorentz
group and Wigner's little groups from the damped harmonic oscillator and
the Poincar\'e sphere.  In addition, it should be noted that the symmetry
of the Lorentz group is also contained in the squeezed state of
light~\cite{knp86} and the $ABCD$ matrix for optical beam
transfers~\cite{bk13mop}.
\par
In addition, we mentioned the possibility of understanding the the mathematics
of the Higgs mechanism in terms of the Lorentz group and its contractions.

\end{document}